\documentclass[twocolumn]{aastex63}
\received{---}
\revised{---}
\accepted{---}
\shorttitle{The GC system of the UDG MATLAS-2019}
\shortauthors{Müller et al.}

\begin{document}

\title{Dwarf galaxies in the MATLAS survey:  Hubble Space Telescope observations of the globular cluster system in the ultra-diffuse galaxy MATLAS-2019}

\correspondingauthor{Oliver M\"uller}
\email{oliver.muller@epfl.ch}

\author[0000-0003-4552-9808]{Oliver M\"uller}
\affiliation{Observatoire Astronomique de Strasbourg  (ObAS),
Universite de Strasbourg - CNRS, UMR 7550 Strasbourg, France}
\affiliation{Institute of Physics, Laboratory of Astrophysics, École Polytechnique Fédéale de Lausanne (EPFL), 1290 Sauverny, Switzerland}

\author{Patrick R. Durrell}
\affiliation{Department of Physics, Astronomy, Geology and Environmental Sciences, Youngstown State University, Youngstown, OH 44555   USA}

\author[0000-0002-1442-2947]{Francine R. Marleau}
\affiliation{Institut f{\"u}r Astro- und Teilchenphysik, Universit{\"a}t Innsbruck, Technikerstra{\ss}e 25/8, Innsbruck, A-6020, Austria}

\author{Pierre-Alain Duc}
\affiliation{Observatoire Astronomique de Strasbourg  (ObAS),
Universite de Strasbourg - CNRS, UMR 7550 Strasbourg, France}

\author{Sungsoon Lim}
\affiliation{University of Tampa, 401 West Kennedy Boulevard, Tampa, FL 33606, USA}

\author[0000-0001-9072-5213]{Lorenzo Posti}
\affiliation{Observatoire Astronomique de Strasbourg  (ObAS),
Universite de Strasbourg - CNRS, UMR 7550 Strasbourg, France}

\author{Adriano Agnello}
\affiliation{DARK, Niels-Bohr Institute, Lyngbyvej 2, Copenhagen, Denmark}

\author[0000-0003-4945-0056]{Rubén Sánchez-Janssen}
\affiliation{UK Astronomy Technology Centre, Royal Observatory, Blackford Hill, Edinburgh, EH9 3HJ, UK}

\author{ M\'elina Poulain}
\affiliation{Institut f{\"u}r Astro- und Teilchenphysik, Universit{\"a}t Innsbruck, Technikerstra{\ss}e 25/8, Innsbruck, A-6020, Austria   }

\author[0000-0002-4033-3841]{Rebecca Habas}
\affiliation{Observatoire Astronomique de Strasbourg  (ObAS),
Universite de Strasbourg - CNRS, UMR 7550 Strasbourg, France}

\author{ Eric Emsellem}
\affiliation{   European Southern Observatory, Karl-Schwarzschild-Str. 2, D-85748 Garching, Germany}
\affiliation{Univ. Lyon, ENS de Lyon, Univ. Lyon 1, CNRS, Centre de Recherche Astrophysique de Lyon, UMR5574, F-69007 Lyon, France }

\author[0000-0003-2922-6866]{Sanjaya Paudel}
\affiliation{Department of Astronomy, Yonsei University, Seoul 03722, Republic of Korea}

\author{Remco F. J. van der Burg}
\affiliation{   European Southern Observatory, Karl-Schwarzschild-Str. 2, D-85748 Garching, Germany}

\author{J\'er\'emy Fensch}
\affiliation{Univ. Lyon, ENS de Lyon, Univ. Lyon 1, CNRS, Centre de Recherche Astrophysique de Lyon, UMR5574, F-69007 Lyon, France }




\begin{abstract}
Ultra-diffuse galaxies (UDGs) are very low-surface brightness galaxies with large effective radii.  Spectroscopic measurements of a few UDGs have revealed a low dark matter content, based on the internal motion of stars or globular clusters (GCs). This is in contrast to the large number of GCs found for these systems, from which it would be expected to correspond to a large dark matter halo mass. Here we present  HST+ACS observations for the UDG MATLAS-2019 in the NGC\,5846 group. Using  the $F606W$ and $F814W$ filters, we trace the GC population two magnitudes below the peak of the GC luminosity function (GCLF). Employing Bayesian considerations, we {identify} 26$\pm$6 GCs associated with the dwarf, yielding a large specific frequency of $S_N=58\pm 14$. {We use the turnover of the GCLF to derive a distance of 21$\pm$2\,Mpc, which is consistent with the NGC\,5846 group of galaxies.}  
Due to the superior image quality of the HST, we are able to resolve the GCs and measure their sizes, which are consistent with the sizes of GCs around Local Group galaxies.
Using the linear relation between the total mass of galaxies and of GCs, we derive a halo mass of $0.9\pm0.2\times10^{11}$\,M$_\odot$ (M$_\odot$/L$_\odot>1000$).  
The high abundance of GCs, together with the small uncertainties, make MATLAS-2019 one of the most extreme UDGs, which likely sets an upper limit of the number of GCs for UDGs.

\end{abstract}


\section{Introduction} \label{sec:intro}
Dwarf galaxies are by far the most numerous galaxies in the Universe \citep{1994AARv...6...67F} and exist in all kinds of environments, from dense clusters to isolated voids (e.g. \citealt{1990A&A...228...42B,2005A&A...437..823R,2015ApJ...808L..39K,2017A&A...597A...7M,2017MNRAS.466..556M}). They are thought to be the most dark matter dominated objects, with the most extreme cases consisting of more than 99\% of this elusive material \citep{2007ApJ...667L..53W,2010ApJ...722..248M,2020MNRAS.491.3496C}. A subsample of the dwarf galaxy population is called ultra-diffuse galaxies (UDGs), a term coined by \citet{2015ApJ...798L..45V}. UDGs are characterized by their large effective radii and very low-surface brightnesses. While their existence has been known since the 1980's \citep[e.g., ][]{1984AJ.....89..919S,1994AJ....107..530M,1997AJ....114..635D,2003AJ....125...66C,2009MNRAS.393..798D,2014ApJ...795L..35C,2014MNRAS.443.3381P}, recently they have become the focus of enhanced attention by the astronomical community, due to the discovery of hundreds of these objects ranging from cluster to group environments \citep[e.g., ][]{2015ApJ...807L...2K,2015ApJ...809L..21M,2016AJ....151...96M,2017ApJ...834...16M,2017A&A...607A..79V,2017A&A...608A.142V,2018A&A...615A.105M,2021A&A...645A..92M,2019ApJS..240....1Z,2020ApJ...899...69L,2020MNRAS.491.1901H,2020A&A...642A..48I}, and as a result  unlocked the potential to study these objects in unprecedented detail.

Due to their diffuse nature, UDGs are an excellent probe of the underlying gravitational potential \citep[e.g., ][]{2019ApJ...874L..12D,2019A&A...627L...1B,2019ApJ...883L..33M,2020MNRAS.495.2582G}. Such studies have been conducted on some specific UDGs, like the two UDGs [KKs2000]\,04/NGC\,1052-DF2 \citep{2000A&AS..145..415K,2018Natur.555..629V} and NGC\,1052-DF4 \citep{2019ApJ...874L...5V} in the NGC\,1052 group of galaxies, NGC\,5846\_UDG1/MATLAS-2019 \citep{2019A&A...626A..66F,2020MNRAS.491.1901H} in the NGC\,5831 group of galaxies, VCC 1287 \citep{1985AJ.....90.1681B,2016ApJ...819L..20B} in the Virgo cluster, and DF17 \citep{2016ApJ...822L..31P} and DF44 \citep{2016ApJ...828L...6V} in the Coma cluster.
The extraction of the stellar velocity dispersion of the body of the galaxy or the velocity dispersion of the globular cluster (GC) population associated to the UDGs has enabled the study of the dark matter content of the galaxies {\citep[e.g., ][]{2018Natur.555..629V,2019ApJ...874L...5V,2019ApJ...884...79C,2019A&A...625A..76E,2020A&A...640A.106M,2020MNRAS.495.2582G}}. While the measurements at face values can be interpreted as an absence of dark matter in these galaxies, both the observational \citep{2018ApJ...859L...5M} and systematic uncertainties \citep{2019MNRAS.484..245L} are too large to yet be conclusive.

Two main mechanisms have been proposed for the formation of UDGs: either they are failed Milky Way like galaxies which couldn't build up their baryonic content \citep[e.g., ][]{2016ApJ...828L...6V,2016ApJ...822L..31P,2018ApJ...856L..31T}, or they {are dwarf galaxies extended to larger radii}, which were puffed up through tidal effects (e.g. tidal interactions or tidal heatings, \citealt{2018MNRAS.480L.106O,2018ApJ...856L..31T,2021MNRAS.tmp...71C}). These two scenarios result in quite different dark matter halo masses for the galaxies. X-Ray observations of multiple UDGs \citep{2019ApJ...879L..12K}, as well as their location in the scaling relations \citep{2020MNRAS.491.1901H}, have revealed that the majority of UDGs are consistent with being normal dwarf galaxies, adding evidence that UDGs are simply the extension of the dwarf galaxy population towards larger radii. Still, this does not exclude the possibility that some UDGs could belong to the former group of failed Milky Way galaxies. One archetypal UDG -- namely DF44 in the Coma cluster -- has been studied in detail with spectroscopy \citep{2016ApJ...828L...6V,2019ApJ...880...91V}, deep imaging \citep{2017ApJ...844L..11V}, and X-ray observations \citep{2020arXiv200907846B}. While initial claims \citep{2016ApJ...828L...6V} pointed it towards the group of failed galaxies -- based on a high count of GCs, as well as an atypical mass for a dwarf galaxy -- these claims have been weakened by more recent  studies \citep{2018MNRAS.475.4235A,2019ApJ...880...91V,2020arXiv200614630S,2020arXiv200907846B}.

One way to {indirectly infer} the mass of a galaxy without measuring the internal stellar dynamics is through the number of {GCs \citep{1997AJ....114..482B,2009MNRAS.392L...1S,2010MNRAS.406.1967G,2014ApJ...787L...5H,2016ApJ...819L..20B,2020AJ....159...56B}.} In a $\Lambda$CDM framework, the dark matter halos of galaxies are directly correlated with the abundance of GCs, i.e. a high number of GCs indicates a vast halo of dark matter. Several studies based on deep imaging have estimated the number of GCs associated to UDGs \citep{2018ApJ...862...82L,2018MNRAS.475.4235A,2019MNRAS.484.4865P}.  They again indicate that UDGs are consistent with being from the dwarf galaxy population, but due to the statistical nature of these studies, the scatter and uncertainties are large. Furthermore, these studies were mainly conducted in dense cluster {environments, and may not be representative of the full UDG population.}

The low-mass environments of groups of galaxies {are} still uncharted territory. This has started to change in recent years \citep[e.g., ][]{2009AJ....137.3009C,2015A&A...583A..79M,2017A&A...597A...7M,2016A&A...588A..89J,2017ApJ...848...19P,2018ApJ...856...69D,2020ApJ...891..144C,2020A&A...644A..91M}. Among these campaigns is the Mass Assembly of early Type gaLAxies with their fine Structures (MATLAS, \citealt{2015MNRAS.446..120D,2020arXiv200713874D,2020MNRAS.498.2138B}) survey, a MegaCam based survey of over 200 nearby early type galaxies within 45\,Mpc. The multi-color imaging in the $ugri$-filters and good image quality made it possible to simultaneously detect dwarf galaxies and GCs. Over 2000 dwarfs have been discovered \citep{2020MNRAS.491.1901H}, with $\sim$5\% of them being UDGs (Marleau et al., accepted). 
The UDG with the highest number of GCs in this systematic survey is MATLAS-2019 (Marleau et al., accepted). This UDG has an effective surface brightness of $\approx25.1$\,mag\,arcsec$^{-2}$ in the $g$ band, an effective radius of 17.2$\arcsec$ ($=2.2$kpc, for the previously assumed distance of $d=26$\,Mpc, corresponding to the center of the NGC\,5846 group), and a systemic velocity of $2156\pm9$\,km\,s$^{-1}$ \citep{2020A&A...640A.106M}, which is consistent with the NGC\,5846 group of galaxies (\citealt{2010A&A...511A..12E}, see also Fig.\,1 in \citealt{2020A&A...640A.106M}).
Spectroscopic follow-up observations with MUSE have confirmed that at least 11 GCs are associated with MATLAS-2019 \citep{2020A&A...640A.106M}, which are all consistent with being metal-poor and old. Intriguingly, at the putative distance to the host group, the brightest GC would be almost as bright as $\omega$ Cen, which is unexpectedly bright for such a low-surface brightness object{. This can also be interpreted as further evidence that $\omega$ Cen is an accreted GC from a bright, now-disrupted dwarf galaxy \citep{2000LIACo..35..619M}}.  However, the data was too shallow to trace the GC population to the faint end. 
A similar case is NGC\,1052-DF2, with the claim that it has a population of too-luminous GCs \citep{2018ApJ...856L..30V}. Either these GCs are indeed too bright, or they are just the tip of the iceberg, i.e. the bright tail of a largely populated GC luminosity function (GCLF). In this work, we present \textit{Hubble Space Telescope} (HST) imaging of MATLAS-2019 to study its GC population by to tracing the full GCLF.

\section{Observations and data reduction} \label{sec:obs}

Our HST images were obtained through a single orbit in the Mid-Cycle 27 program GO-16082, (PI: M\"uller) to observe the UDG MATLAS-2019. {This galaxy was first identified in \citet{2005AJ....130.1502M} and named N5846-156, and has been identified} in the MATLAS survey \citep{2020MNRAS.491.1901H}, as well as independently reported in the VST Early-type GAlaxy Survey (VEGAS, \citealt{2019A&A...626A..66F}). 
We observed MATLAS-2019 with the HST Advanced Camera for Surveys (ACS) in the F606W and F814W filters.  The galaxy itself was placed at the center of one of the CCDs to maximize a suitable background/control sample.   Two dithered images (separated by 0.5$\arcsec$) of 515\,s each were taken in each filter. We used the final, reduced  CTE-corrected (Charge Transfer Efficiency corrected) {\it .drc.fits} images produced by the standard pipeline and used the VEGAmag zeropoints provided by the ACS online documentation\footnote{https://www.stsci.edu/hst/instrumentation/acs/data-analysis/zeropoints.}.

\begin{figure*}[ht!]
\plotone{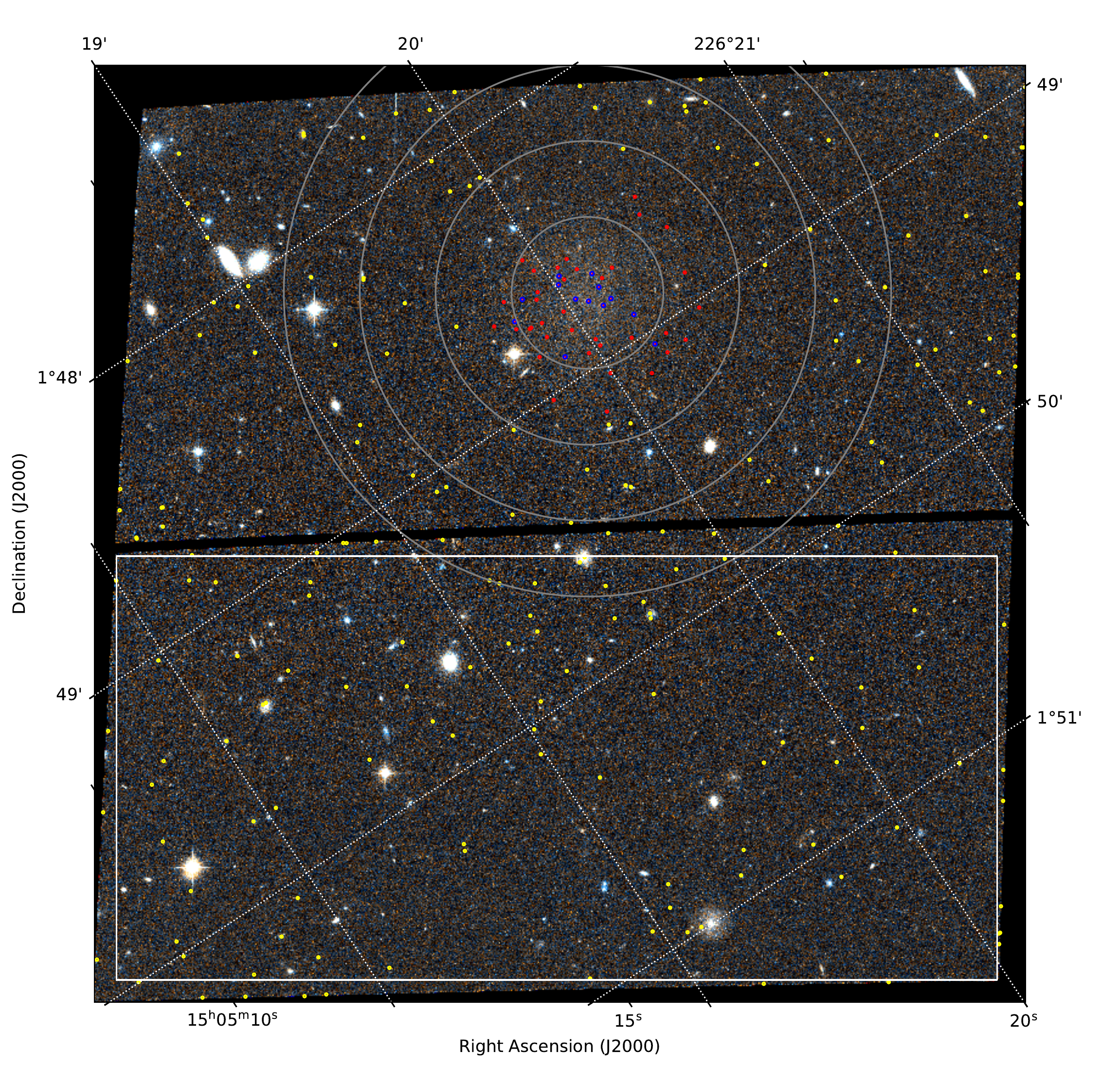}
\caption{The full ACS field of view of our HST observations. In red we indicate the detected GC candidates, in blue the confirmed GCs from MUSE observations, and in yellow we show the selected field GC candidates. The gray cirles are multiples of the effective radius, starting with 1\,$r_{eff}$. The white box indicates the reference field.  \label{fig:field}}
\end{figure*}
\begin{figure*}[ht!]
\plotone{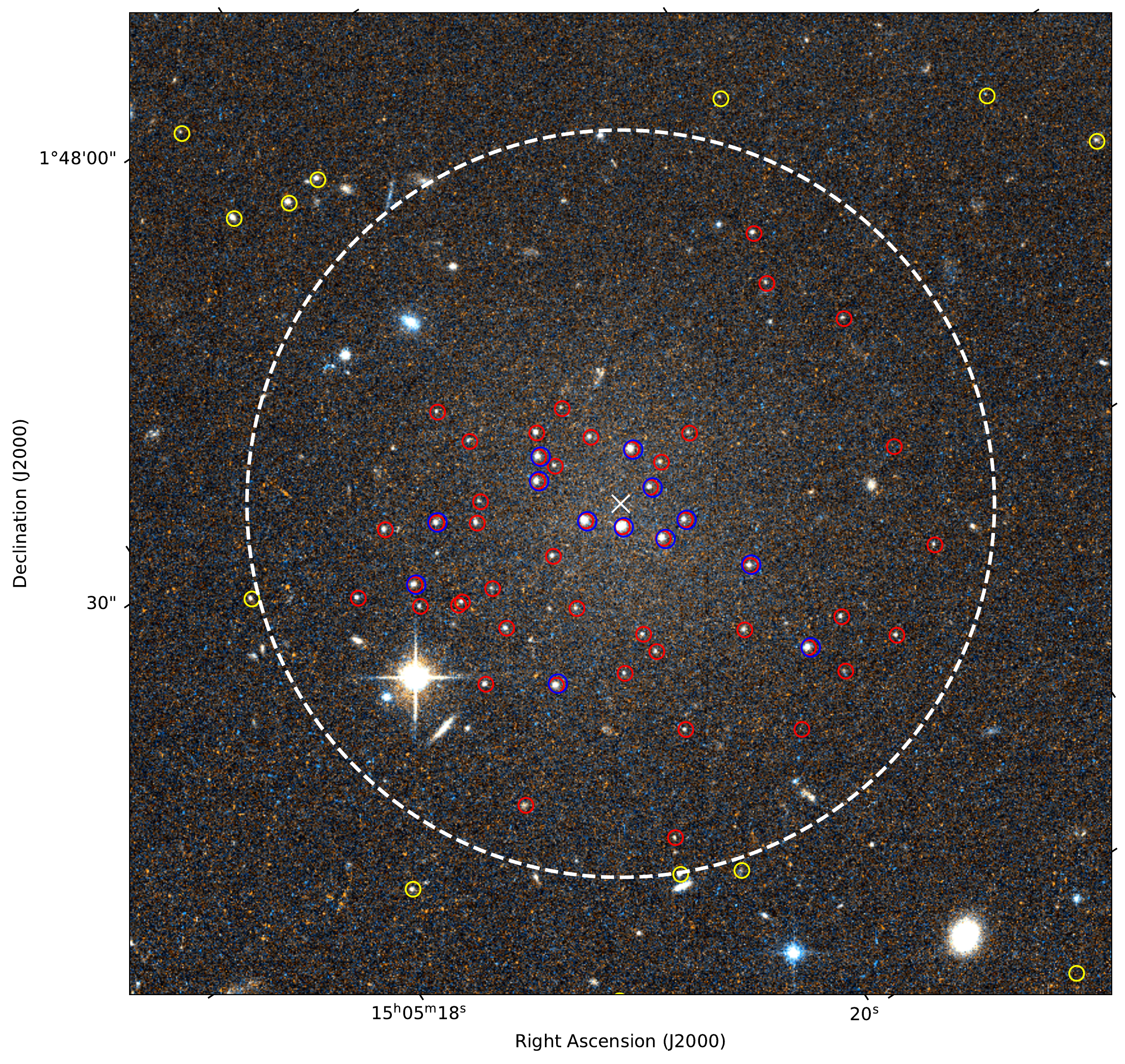}
\caption{The zoom in view of our HST observations. {The dashed white line corresponds to the radius where the GC candidate count drops to a background level (i.e. at 1.75\,$r_{eff}$), corresponding to our selection radius. The white cross indicates the galaxy center.}
In red we indicate the detected GC candidates, in blue the confirmed GCs from MUSE observations, and in yellow we show the selected field GC candidates. One side is $\approx$80$\arcsec$.  \label{fig:zoom}}
\end{figure*}

We performed aperture photometry on each filter with the python package photutils \citep{larry_bradley_2020_4044744}. To remove the smooth light profile of the galaxy, we have modelled and subtracted it with \textsc{galfit} \citep{2002AJ....124..266P}. {The center of the galaxy is estimated to be at 15:05:20.34, $+$01:48:44.9.}
Then, we ran sep \citep{SEP}, the Source Extractor  \citep{1996A&AS..117..393B} implementation in python, with a 3$\sigma$ threshold and a minimum of 5 adjacent pixels, to create a catalog of objects. On this catalog we made a first cut, namely we rejected all sources with a detected radius smaller than 3\,px or an ellipticity larger than 0.2. This removed noise peaks as well as elongated objects such as background galaxies. We performed photometry on the detections in the catalog with a circular aperture with a radius of 3\,px. The median background value was estimated in an annulus with inner and outer radii of 8\,px and 23\,px, respectively, employing a 3$\sigma$ clipping, and was subtracted from the photometry of the object. Aperture corrections were applied according to the ACS documentation\footnote{https://www.stsci.edu/hst/instrumentation/acs/data-analysis/aperture-corrections.}. The $F606W$ and $F814W$ magnitudes were corrected for extinction ($A_{F606W}=0.131$\,mag and $A_{F814W}=0.081$\,mag) according to the extinction calculator tool on NED\footnote{http://ned.ipac.caltech.edu/extinction\_calculator.} (using \citealt{2011ApJ...737..103S}).
We then matched the catalogs in the two filters with a 0.5\,arcsec tolerance. Finally, we applied a magnitude cut in the $F606W$ band of $22.0<F606W<26.2$\,mag, as well as a $(F606W-F814W)_0$ color cut of $0.5<(F606W-F814W)_0<0.9$\,mag, motivated by the range of expected magnitudes and colors of GCs.  From this we generated a catalog of 241 GC-like sources over the entire ACS FOV.   Of these, we extract a sample of $N_{tot}=49$ GC candidates located within 1.75\,$r_{eff}$ of the galaxy center (this optimal aperture size is derived in Section\,\ref{sec:radial}).
The GC candidates are presented in Fig.\,\ref{fig:field} and Fig.\,\ref{fig:zoom}.
 {As a reference background field, we use  an area of $100<x<4090$ and $100<y<2020$ pixels on the second chip, containing 74 sources in our catalog. This area is well outside of 3\,$r_{eff}$ (see Fig.\,\ref{fig:field}).}

\begin{figure}[ht!]
\plotone{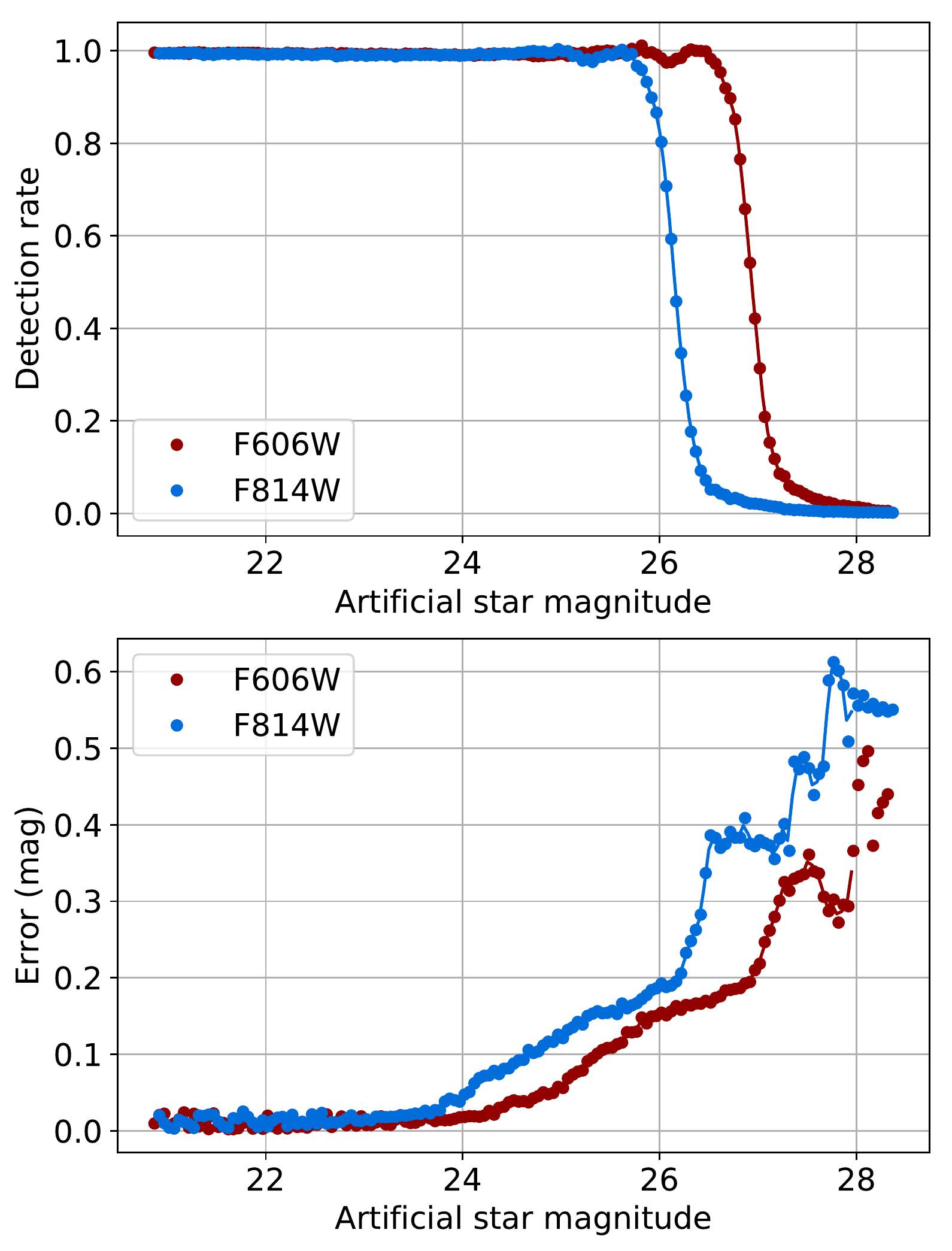}
\caption{Results of the artificial star experiments. The top panel shows the detection rate, the bottom panel the mean uncertainty of the recovered artificial stars. \label{fig:completeness}}
\end{figure}

To assess our completeness and photometric errors we have injected artificial point sources in the HST data and re-run our detection and photometry pipeline. For that we have created a PSF model using 14 {bright, isolated stars on the frame.} Artificial stars were added on a evenly-spaced grid {with a separation of 70 pixels. This gives $\approx$2500 injected objects}. We have repeated this for a magnitude range between 21 and 28.3 in both filters. To obtain the completeness, we have compared the number of injected vs. number of detected objects. In order to  not count real objects as successfully detected artificial objects, we have provided Source Extractor with the original segmentation map of the image as a mask. For the photometric error we have compared the injected and extracted magnitudes. The completeness drops below 95\% at $F606W=26.74$\,mag and $F814W=25.82$\,mag. The results of our artificial star experiments are shown in Fig.\,\ref{fig:completeness}. 
{To test whether the light from MATLAS-2019 impacts the performance we consider the detections within one effective radius of the galaxy separately. Within this radius, $\approx$70 objects are injected per iteration. From these we can again construct a completeness curve. Apart from the the curve being more noisy -- due to having less objects to work with -- we find no difference in the shape and the reached completeness limit.}

To convert the HST $F606W$ and $F814W$ magnitudes into the standard $BVI$ system, we employed the transformations provided by \citet{2018AJ....156..296H} and assumed a $(V-I)_0=0.9$\,mag color for the GCs.

\begin{equation}
V=F606W+0.15(V-I)_0
\end{equation}

\begin{equation}
I=F814W-0.13+0.108(V-I)_0
\end{equation}

{Due to the superior image quality of the HST the GCs in our data are partially resolved. To measure their half light sizes ($r_h$) {, core sizes ($r_c$) and tidal radiis ($r_t$)}, we have employed \textsc{galfit} using a S\'ersic profile {and a King profile, respectively, which were} convolved with the PSF model.} {For the fitting, we let the S\'ersic index be a free parameter {smaller than five}, which helped reduce crashes. All photometric properties are compiled in Table\,\ref{tab:sample}.} 

\section{Properties of the Globular Cluster population}

\subsection{Globular Cluster Luminosity Function}

In the following we will model the GCLF using Bayesian considerations. For the GCLF we assume a {log-}normal distribution $\mathcal{G}_V$ {in luminosity}:

\begin{equation}
   \mathcal{G}_{V,j}(m_V,\delta_{V,err})=
    \frac{1}{\sqrt{2\pi}}\frac{1}{\sigma_{obs}} \exp{\frac{-(m_{V,j}-\mu_V)^2}{2\sigma_{obs}^2}},
\end{equation}
with
\begin{equation}
 \sigma_{obs}=\sigma_{V,j}^2 + \delta_{V,err}^2.
\end{equation}
The magnitude of the GC candidate $j$ is given by $m_{V,j}$. The total width $\sigma_{obs}$ is the combination of the intrinsic width of the GCLF  $\sigma_V$ and the mean error $\delta_{V,err}$ of all the GCs. 
To model the background contamination $\mathcal{B}_V$ we use all GC-like  objects outside 1.75 $r_{eff}$ to create a histogram with a bin size of 0.5\,mag. We then interpolate the midpoints of the bins to create $\mathcal{B}_V$.
Similarly, we model  the color of the GC distribution with a normal distribution $\mathcal{G}_{(V-I)_0}$ and derive the background contamination $\mathcal{B}_{(V-I)_0}$ from all the GC-like objects outside 1.75\,$r_{eff}$. {We assume  a normal distribution for the colors of the GCs of MATLAS-2019 because when we inspected the color magnitude diagram (CMD, see Sec\,\ref{subsec:colors}), we found that the color distribution is strikingly unimodal. In general, this doesn't need to be the case.}

We use a Plummer profile $\mathcal{P}$ to model the spatial distribution of the GCs:

\begin{equation}
   \mathcal{P}(r)_j =  \frac{1}{\pi} \frac{1}{r_{GC}^2(1+r_j^2/r^2_{GC})^2},
\end{equation}
where $r_j$ is the galactocentric distance of the GC candidate j and $r_{GC}$ is the half number radius of the GC system, which is a free parameter.
To be consistent in the dimensions, we multiply this density with the number of GCs $N_{GC}$ associated with the galaxy:

\begin{equation}
\Sigma(r)_j = N_{GC}\cdot\mathcal{P}(r)_j 
\end{equation}

The contamination $c$ out to the outermost GC candidate (having the radius $R_{out}$) is given by:
\begin{equation}
c = \frac{N-N_{GC}}{\pi R_{out}^{2}}.
\end{equation}

Finally, the likelihood for a source $j$ in our catalog is given by:

\begin{equation}
L_j= \frac{\Sigma_j \cdot \mathcal{G}_{V,j}  \mathcal{G}_{(V-I)_0,j}+ c\cdot\mathcal{B}_{V,j}\mathcal{B}_{(V-I)_0,j}} {\Sigma_j+ c}.
\end{equation}

\begin{table}[ht]
\caption{Priors for our Markov Chain Monte Carlo. 
}             
\centering                          
\begin{tabular}{l c}        
\hline\hline                 
Parameter & Prior \\
\hline      \\[-2mm]                  
 $\sigma_V$ & 0 $<$  $\sigma_V$ $<$ 2 \\
  $\mu_V$ & 22 $<$  $\mu_V$ $<$ 26.8 \\
    $N_{GC}$ & 0 $<$  $N_{GC}$ $<$ 49 \\
    $R_{out}$ & 0 $<$  $R_{out}$ $<$ 30 \\
$(V-I)_0$ & 0.6 $<$  $(V-I)_0$ $<$ 1.1 \\
$\sigma_{(V-I)_0}$ & 0.0 $<$  $\sigma_{(V-I)_0}$  \\
\hline
\end{tabular}
\label{tab:prior}
\end{table}

 The parameters we are marginalizing over are $\mu_V$ and $\sigma_V$, i.e.  the peak and the width of the GCLF; the number of true GCs $N_{GC}$ associated with MATLAS-2019; the half number radius $r_{GC}$ of the GC system; and $\mu_{(V-I)_0}$ and $\sigma_{(V-I)_0}$, i.e.  the peak and the width of the color distribution. As priors  we use flat priors, namely for  $\sigma_V$ we allow values between 0 and 2\,mag, which are well motivated by observations \citep{2012Ap&SS.341..195R}; for $\mu$ we impose the condition that it must be within our magnitude range (i.e being between 22 and 26.8\,mag); for $N_{GC}$ we impose that the number must be between 0 and the number of detected sources within our aperture ($N=49$); for $r_{GC}$ we set them to be between 0 and $R_{out}$ (i.e. 30$\arcsec$); and for the colors we demand that they must be between $0.6<(V-I)_0<1.1$\,mag
 and that $\sigma_{(V-I)_0}$ must be larger than 0. {A summary of the prior constraints is given in Table\,\ref{tab:prior}.}

\begin{figure*}[ht!]
\plotone{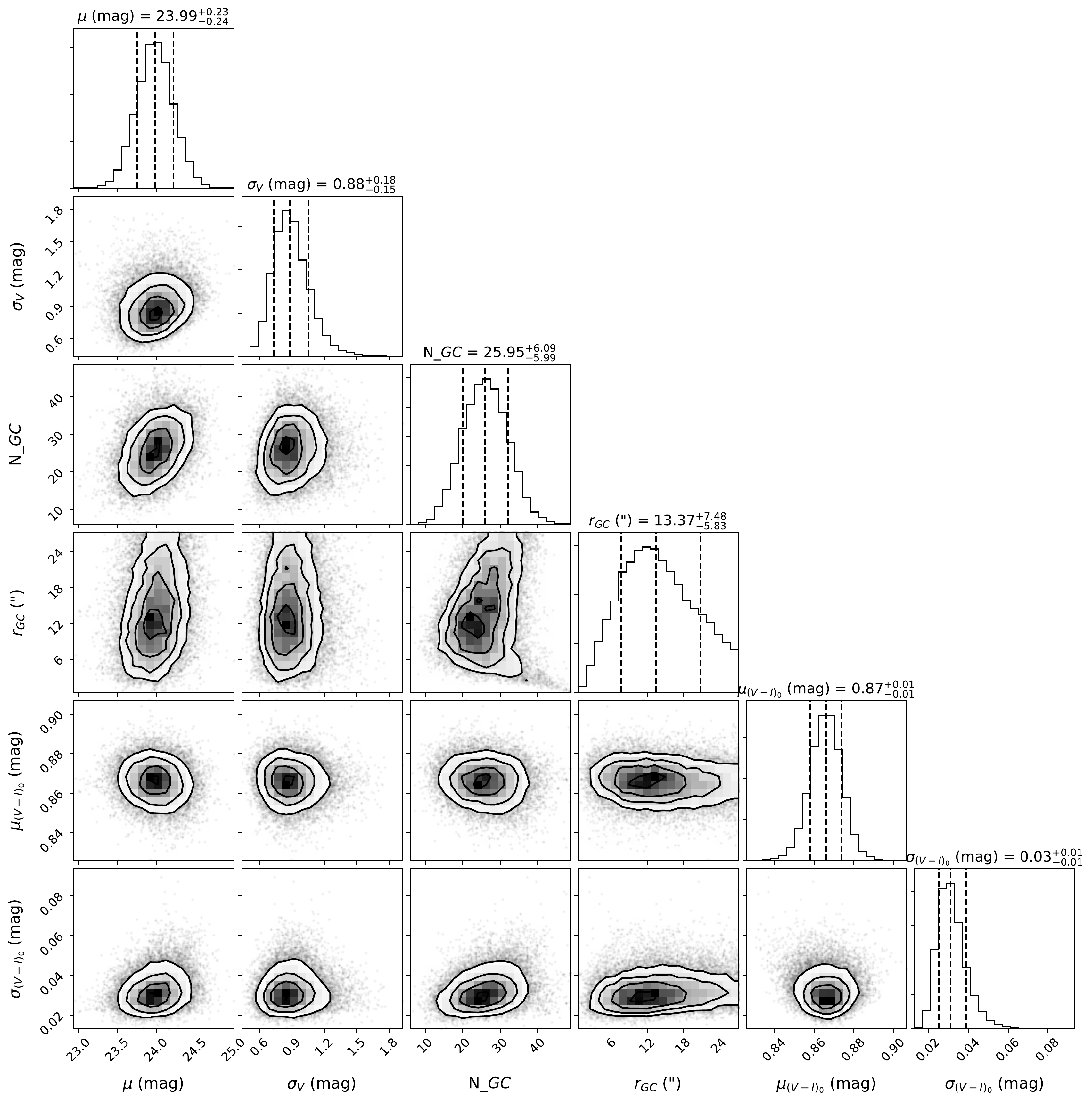}
\caption{
The posterior distributions from our MCMC modelling with its six free parameters  ($\mu_V$, $\sigma_V$, $N_{GC}$,  $r_{GC}$, $\mu_{(V-I)_0}$, and $\sigma_{(V-I)_0}$). \label{fig:posterior}}
\end{figure*}

To estimate the posterior distributions for these six parameters ($\mu_V$, $\sigma_V$, $N_{GC}$,  $r_{GC}$, $\mu_{(V-I)_0}$, and $\sigma_{(V-I)_0}$) we employ a Markov Chain Monte Carlo (MCMC) algorithm \citep{2010CAMCS...5...65G} implemented through the python package emcee \citep{2013PASP..125..306F}. As a first guess for the parameters, we use a maximum likelihood method implemented in scipy (i.e. minimize from the scipy.optimize module), which yields $\mu_V=23.99$\,mag, $\sigma_V=0.81$\,mag, $N_{GC}=24.2$, and $r_{GC}=13.7$\,$\arcsec$, $\mu_{(V-I)_0}=0.87$\,mag, and $\mu_{(V-I)_0}=0.03$\,mag.
For the MCMC we use 100 walkers with 10 000 steps along the chains, with a burn-in of 1000 steps. For the peak of the GCLF we derive $\mu_{V}=23.99\pm 0.24$\,mag, for the width of the GCLF $\sigma_{V}=0.88\pm0.16$\,mag, for the number of GCs $N_{GC}=26.0\pm6.0$, for the half number radius $r_{GC}=13.4^{+7.5}_{-5.8}\arcsec$, and for the color a mean of $(V-I)_0=0.87\pm0.01$\,mag with a spread of $\sigma_{(V-I)_0}=0.03\pm0.01$\,mag. 
The posteriors are well-behaved and presented in Fig\,\ref{fig:posterior}.

\begin{figure*}[ht!]
\plottwo{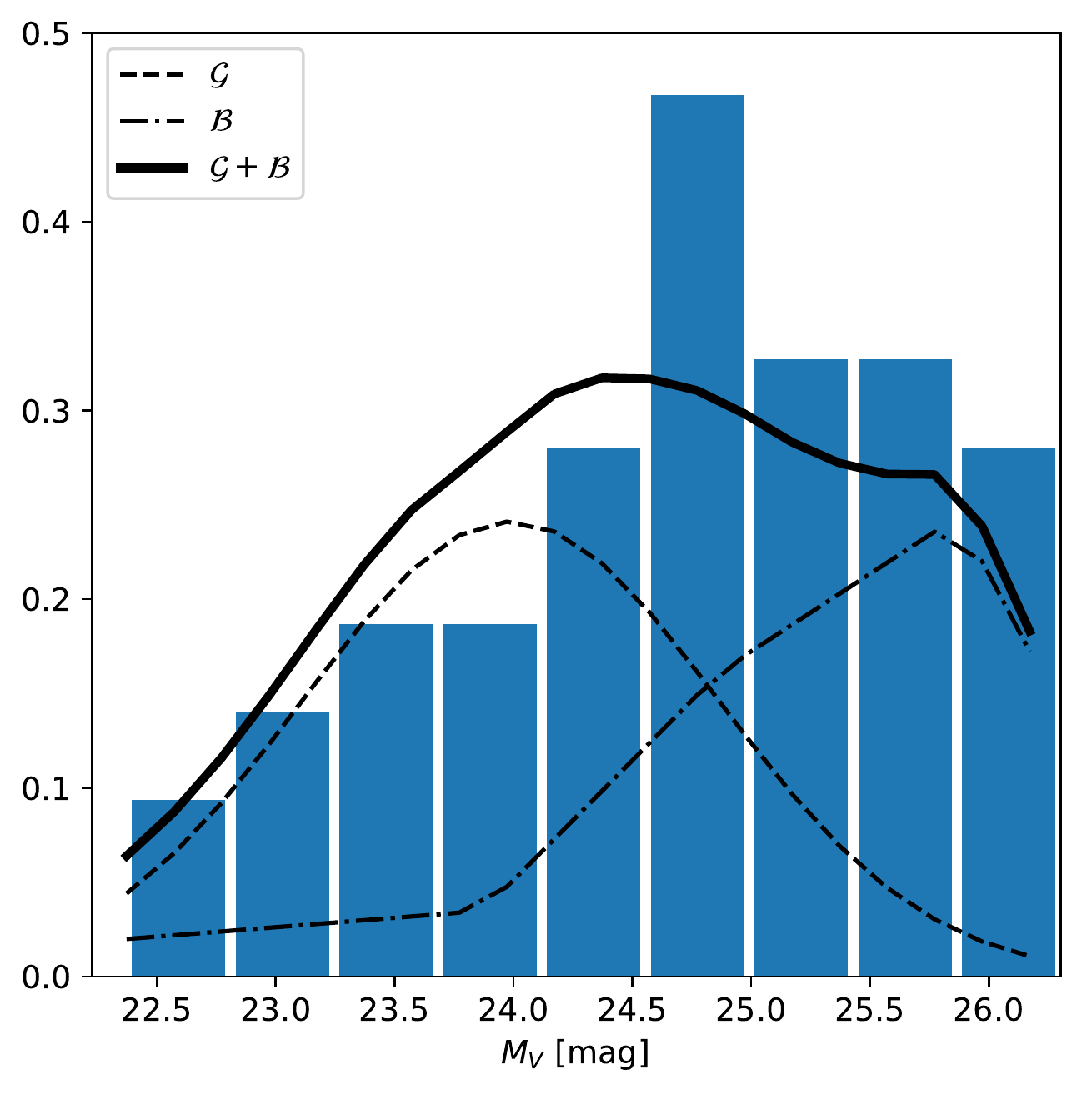}{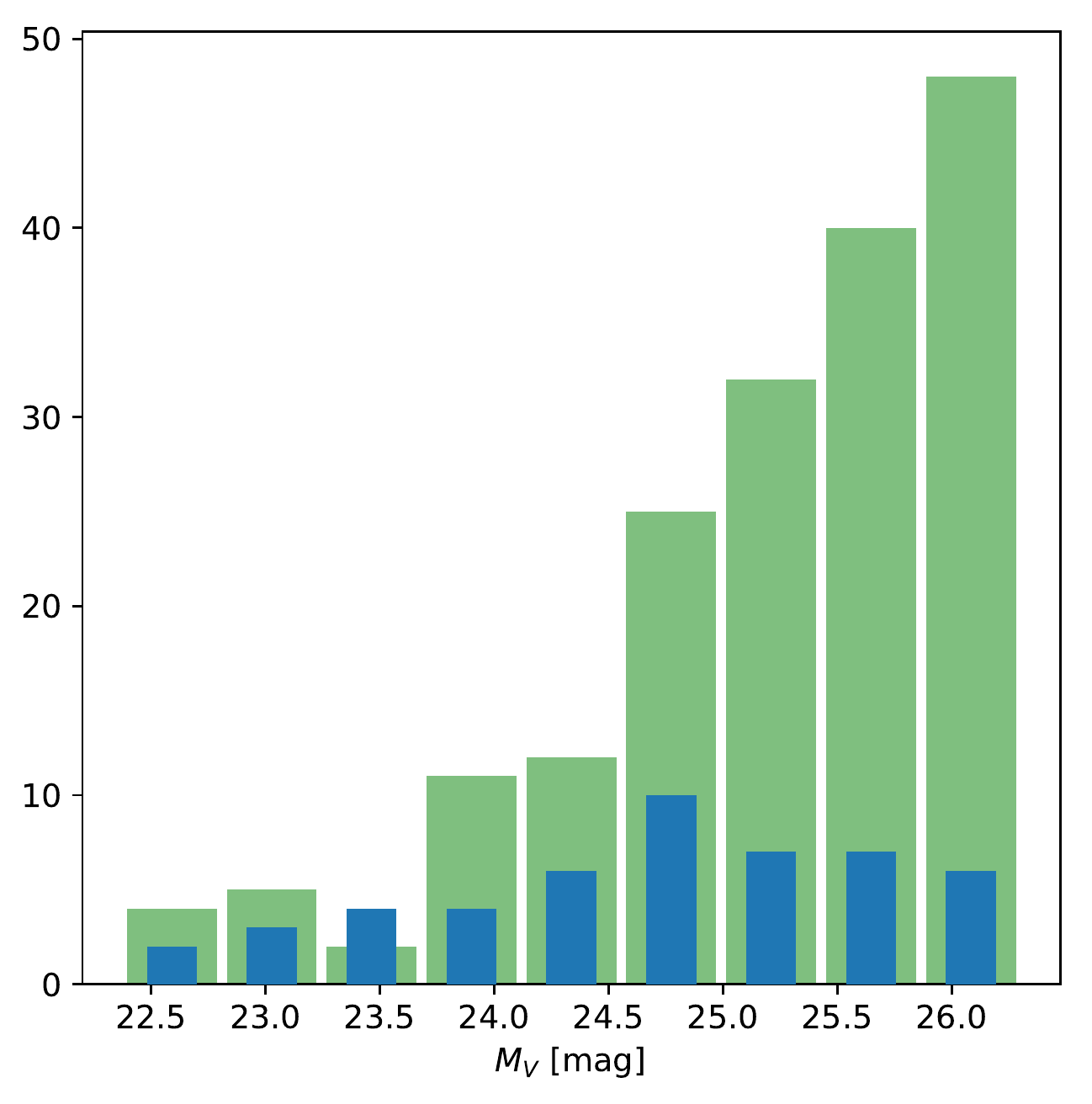}
\caption{
Left: The {normalized} histogram of the $V_0$ band magnitudes of the {GC candidates within 1.75\,$r_{eff}$}. The dashed line corresponds to the GCLF, the dash-dotted line to the background model, and the thick line to the combination of the two. {Right: The  histogram of the $V_0$ band magnitudes of GC candidates within the 1.75\,$r_{eff}$ aperture (blue) and outside of this aperture (green).}
\label{fig:gclf}}
\end{figure*}

The histogram of the GC candidates within 1.75 $r_{eff}$ and the best-fitting model are presented in Fig.\,\ref{fig:gclf}. The peak of the GCLF puts the galaxy at $20.6_{-2.1}^{+2.3}$\,Mpc (assuming the peak of the GCLF is at $-$7.6$\pm$0.1\,mag, \citealt{2012Ap&SS.341..195R}). This is consistent with the distance of the NGC\,5846 group (having a mean of 26\,Mpc, \citealt{2011MNRAS.413..813C}, but having { giant galaxies within} a spread of $\approx\pm$5\,Mpc). Our MCMC modelling reveals a large number of GCs for such a low-luminosity dwarf. The brightest GCs are consistent with the normal bright end of the GCLF at the distance of the NGC\,5846 group,  indicting there are no over-luminous GCs present in this galaxy, as has been suggested \citep{2020A&A...640A.106M}. Rather, the GCLF is best explained by a large population of GCs for a galaxy of this luminosity. {For reference, in Fig.\,\ref{fig:gclf} we further show the histogram of the detected GC candidates within and outside of the galactic aperture. }

As a check, we estimate the numbers of GCs expected in our aperture of the size 1.75 $r_{eff}$, when considering the contamination estimated on our reference background field. With an area of 5.3\,arcmin and 74\,GC-like sources in this reference background area, we expect $\approx$13 GC candidates in our aperture to be interlopers. In other words, there should be an over-density $36\pm6$ detection within our aperture {assuming Poisonnian noise for the estimation of the uncertainty}. {This number is 
larger than the number of GCs ($N_{GC}=26\pm6$) estimated by our MCMC scheme. This can be explained by Poisson noise: the mode or median value of a Poisson distribution is lower than its expectation value, meaning that when we estimate the contamination from our background field we will likely underestimate the number of contaminants  (i.e. overestimate the number of GCs associated with the dwarf). However, within the uncertainties, these numbers are still consistent.}

{To test that this difference between our MCMC approach and the overdensity estimate of the number of GCs is not due to our model, we produced a synthetic data set where we had the control over the number of GCs and the other parameters of interest. For this purpose, we used the same number of detections as for our analysis. To create the background, we used our background luminosity and color model together with a random position to draw the data points. For our synthetic GCs, we drew their luminosities from a gaussian with the peak and standard deviation as our estimated values ($\mu=24$\,mag and $\sigma=0.9$\,mag), with a color of $(V-I)=0.9$\,mag, and a positions drawn from the Plummer profile with the estimated value of $r_{GC}=12.6$\,arcsec. All the remaining inputs were kept the same. With this setup, we were able to recover all input parameters within their uncertainties, with slightly overestimating the number of GCs (on average getting $28.9\pm2.2$ GCs instead of 26).}

\subsection{Luminosities{, colors, and sizes}}
\label{subsec:colors}
\begin{figure*}[ht!]
\plotone{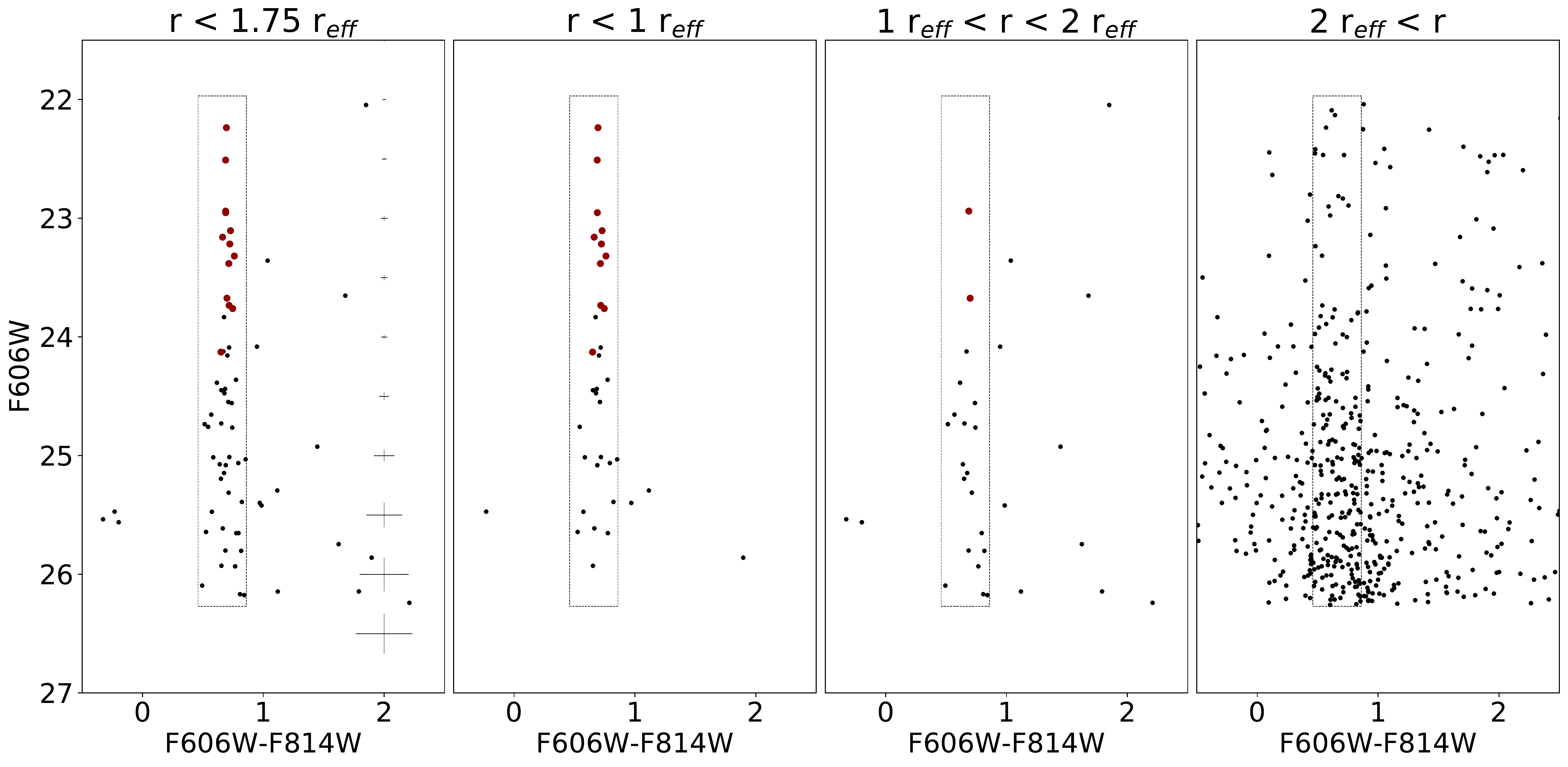}
\caption{The CMD of detected sources within 1.75 $r_{eff}$ (left), corresponding to the full extent of the GC system; within 1 $r_{eff}$ (middle left);  within 1 and 2 $r_{eff}$ (middle right); and outside of 2 (right). The dotted line corresponds to our selection of GCs. In the left panel, we indicate in red the spectroscopically confirmed GCs from \citet{2020A&A...640A.106M}. \label{fig:CMD}}
\end{figure*}

In Fig.\,\ref{fig:CMD} we present the CMDs of our observations. We considered three cases:  all detected point sources within  i) one effective radius ($r_{eff}$), ii) between 1 and 2\,$r_{eff}$, and iii) outside of 2\,$r_{eff}$. Remarkably, within 1\,$r_{eff}$ the GCs follow an extremely tight sequence. All confirmed GCs from our MUSE observations \citep{2020A&A...640A.106M} within 1\,$r_{eff}$ are located within this sequence. For the other two CMDs, the color scatter of the GCs increases, which is due to an increase of the background contamination. This indicates that the GC population contribution drops after one effective radius and the contamination by background/foreground sources increases. This is a similar trend as seen the GC population of DF44, where the GC population drops significantly after one effective radius \citep{2020arXiv200614630S}. We will quantify this further in the next subsection.

 {For most GC candidates, the core radius is systematically smaller than the half-light radius, as has been found for Cen\,A GCs \citep{2002AJ....124.1435H}. As a sanity check, the fraction  between the tidal radius and the core radius under a logarithm with the base of 10 should roughly be between 1 and 2 for confirmed GCs \citep{2002AJ....124.1435H}. This is the case for most of the GC candidates here, especially within the given large errors of these difficult observations. The sizes are not used in the selection for the GC candidacy though, making them an independent property. This suggests that most of the GC candidates selected by their magnitudes and colors are also GC-like in their morphology and are therefore reasonable GC candidates. }

\begin{figure*}[ht!]
\plottwo{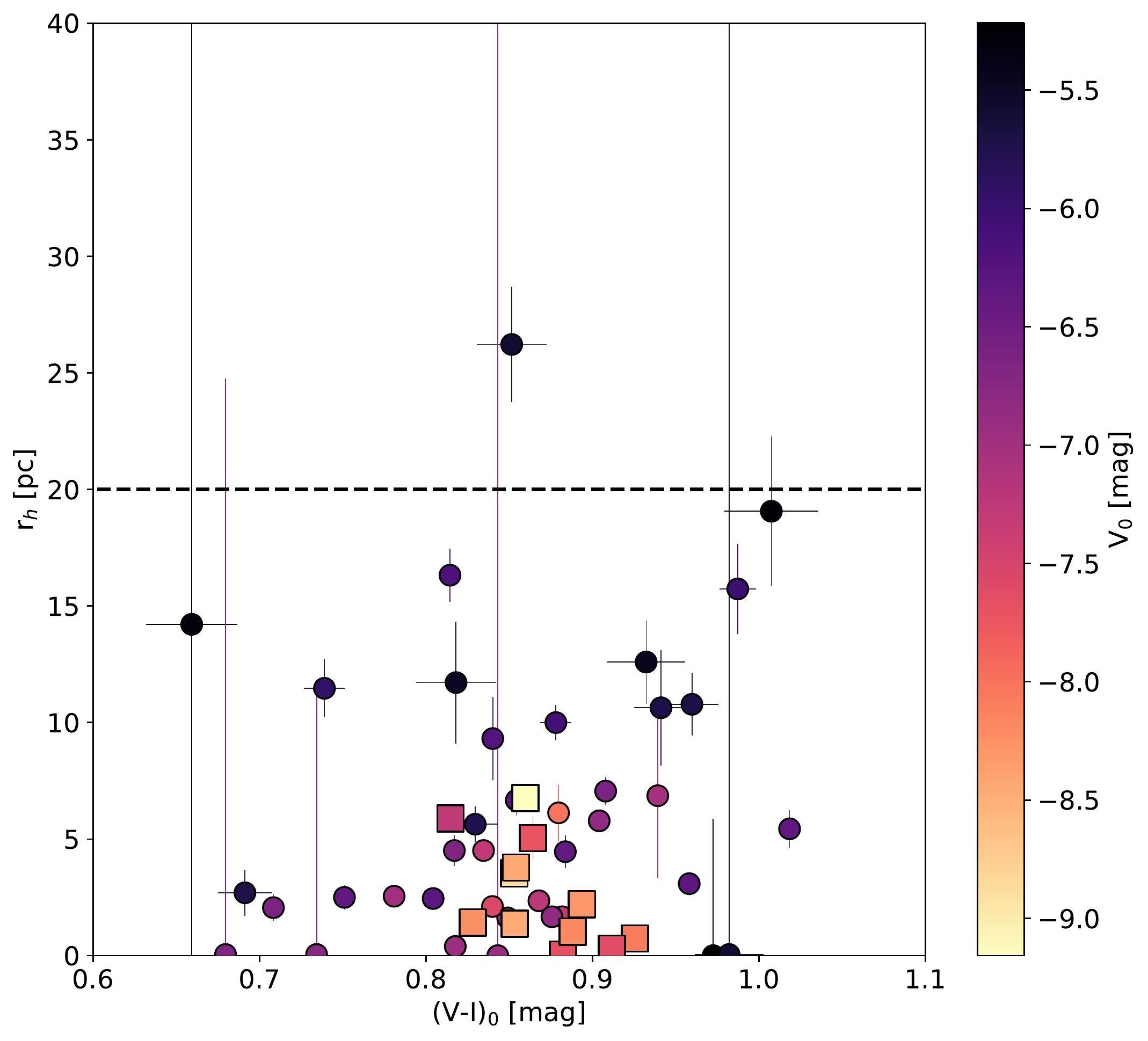}{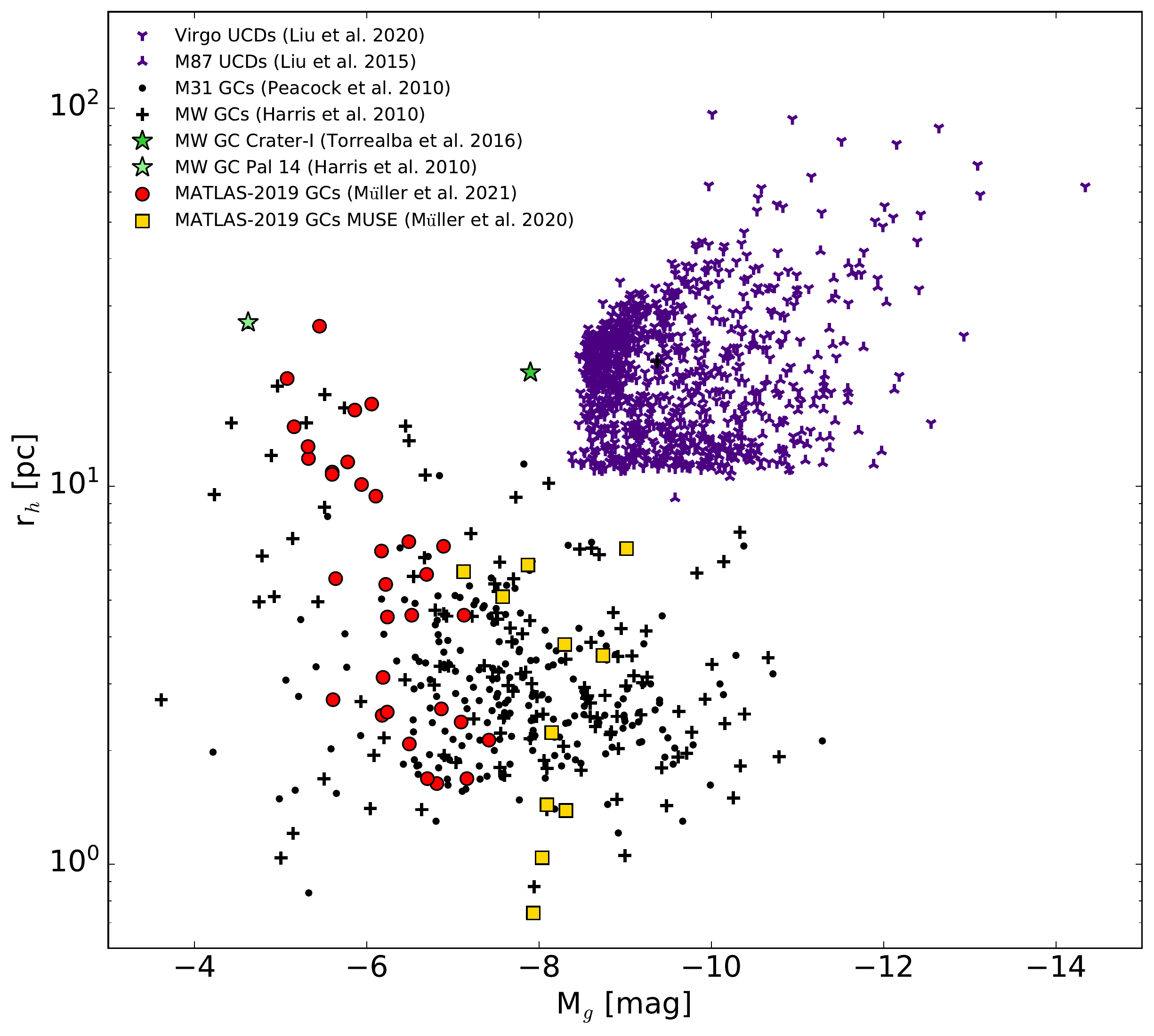}
\caption{Left: The size of the GCs as a function of their $(V-I)_0$ color. The points are further shaded based on their luminosity. The spectroscopically confirmed GCs \citep{2020A&A...640A.106M} are represented with the squares. The dashed line indicates the size of Crater-I \citep{2016MNRAS.459.2370T}, an extreme example of an extended GC in the Milky Way. Right: The size of the GCs as a function of their luminosity and compared to Local Group GCs \citep{2010MNRAS.402..803P,2010arXiv1012.3224H} and Virgo cluster UCDs \citep{2015ApJ...812...34L,2020ApJS..250...17L}. The red filled circles indicate the GCs associated with MATLAS-2019 and the gold filled squares are those spectroscopically confirmed with MUSE. \label{fig:color_size}}
\end{figure*}

In Fig.\,\ref{fig:color_size} (left panel) we plot the sizes (r$_h$) of the GCs as functions of luminosity and color.  Most of the GC candidates are smaller than 10\,pc, which is well expected from studies of the sizes of Milky Way GCs \citep{2010arXiv1012.3224H}.   All of our spectroscopically confirmed GCs are in this range.   Some of the faint GC candidates, however, have larger sizes than the brighter GCs.  This could have a few explanations.   The fainter GC candidates have a lower S/N, which could systematically increase their sizes due to modelling uncertainties (which are not well captured by galfit). Alternatively, and more likely, they are the background objects contaminating our selection.  We expect $N_{tot}-N_{GC}=23$ objects in our catalog to be background interlopers.   It is also possible a few of these objects could be faint, diffuse GCs, much like those seen in M31 \citep{2014MNRAS.442.2165H} with $r_h$ values up to 30-35 pc.  Up to five GC candidates (including their uncertainties) could have half light radii larger than the one of Crater-I (i.e. $r_{eff}=20$\,pc, \citealt{2016MNRAS.459.2370T}) -- which is one of the most extreme GCs in terms of size associated with the Milky Way \citep{2016MNRAS.460.3384V} -- and could therefore be such extreme GCs. However, until spectroscopically confirmed, {it is more likely that} those are the expected interlopers. 

In Fig.\,\ref{fig:color_size} (right panel) we also plot the GC candidates of MATLAS-2019 compared to other known GCs from the Local Group \citep{2010MNRAS.402..803P,2010arXiv1012.3224H}, as well as ultra-compact dwarfs (UCDs, \citealt{1999A&AS..134...75H}) in the Virgo cluster \citep{2015ApJ...812...34L,2020ApJS..250...17L}. The particular
cases of Crater-I and Pal 14, the largest Milky Way GC,
are highlighted in the figure. For the UCD FCC\,47-UCD1 \citep{2019A&A...625A..50F} in the Fornax cluster,  \citet{2019A&A...625A..50F} (among others) have suggested that it could originate from the brightest star cluster in the progenitor galaxy and was accreted through a minor merger, {and it could be interesting to see if we would have such a compact stellar object in our sample}. 
 While most of the GC candidates associated with MATLAS-2019 follow the distribution of the Local Group GCs, one GC in particular -- the brightest GC in our sample dubbed GC6 in \citet{2020A&A...640A.106M} -- is approaching the UCD regime, with a size of $r_{eff}=6.7\pm0.2$\,pc and an absolute magnitude of $-9.2$\,mag in the $V$ band. In Table\,\ref{tab:sample}, GC6 is number 29. We note that the GCLF peak distance prefers a value of 21\,Mpc, while the velocity of the galaxy rather puts it at 25 to 30\,Mpc. If the dwarf is at a larger distance, the sizes and luminosities of the GCs would increase, and GC6 would be closer to the UCD regime ($r_{eff}\approx10$\,pc and $M_V\approx-10$\,mag). However, GC6 would fall out of the selection of UCDs in the Virgo cluster { -- independent of whether the dwarf is at 21 or 30\,Mpc --}, with the GC being too small to be considered a UCD. {So with their selection criteria, GC6 is not a UCD. \citealt{2019A&A...626A..66F} suggested that this GC is a nuclear star cluster, based on its brightness and central placement within the galaxy. Its velocity is similar to that of the stellar body \citep{2020A&A...640A.106M,2021MNRAS.500.1279F}, further supporting this assessment.}

\subsection{Radial profile}
\label{sec:radial}
\begin{figure}[ht!]
\plotone{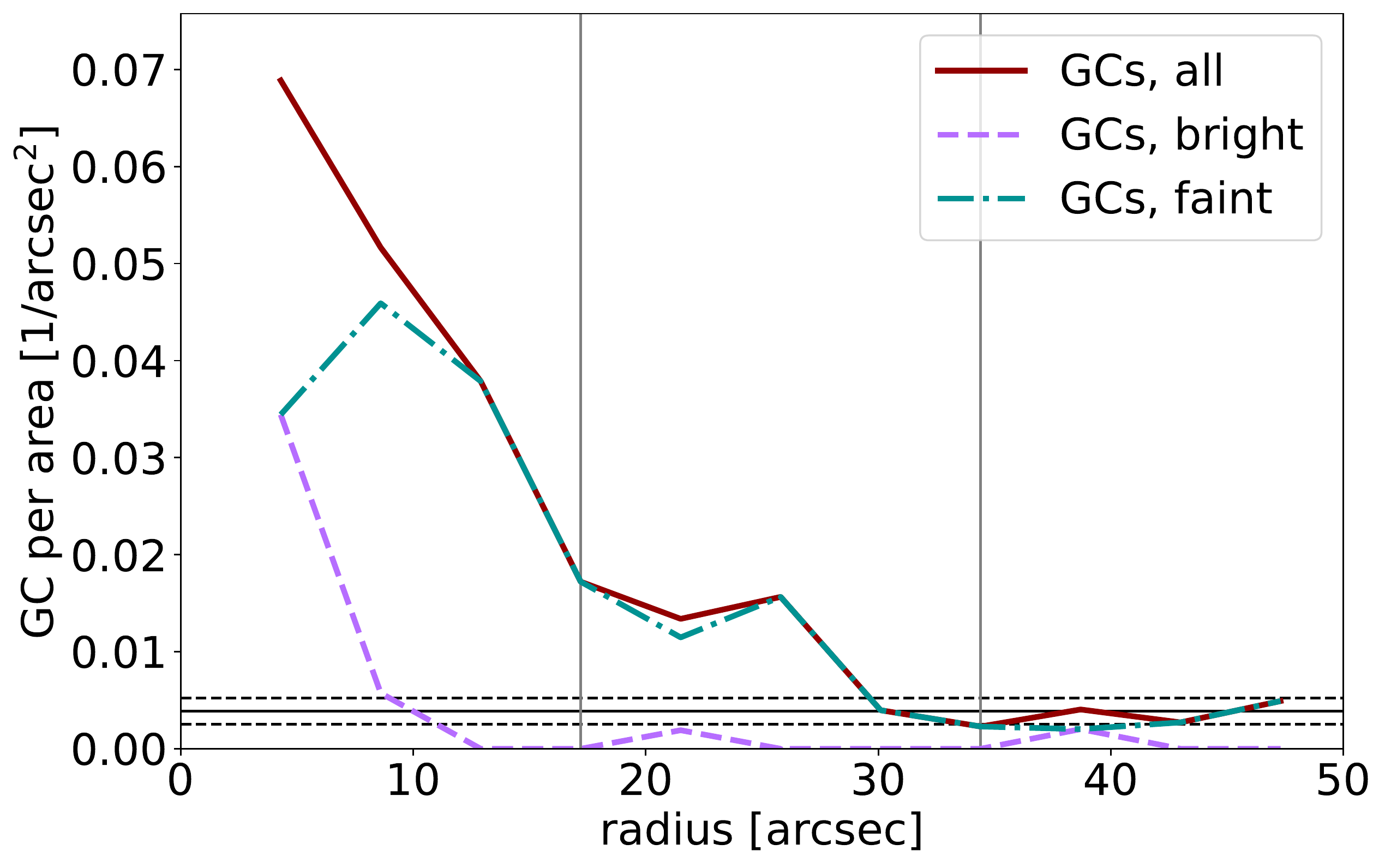}
\caption{The GC density (red line) in increasing annuli of radii of 0.25\,$r_{eff}$. The densities of bright GCs and fainter GCs are marked with the violet dashed line and cyan dash-dotted line, respectively.  The vertical lines indicate one and two effective radii, respectively. The horizontal straight and dashed black lines represents to the background GC density on chip two of the ACS field.  \label{fig:radius}}
\end{figure}

In the following, we quantify the radial extent of the GC distribution. For this, we counted the number of GCs per area in increasing annuli of multiples of 0.25\,$r_{eff}$. The results are presented in Fig.\,\ref{fig:radius}. At 1\,$r_{eff}$, the GC density is at roughly 10\% compared to the innermost circle, and after  1.75\,$r_{eff}$, the GC count drops to the background level (being calculated from the density of GC candidates on the reference field on the second chip). While most of the GCs reside within one effective radius, the GC population extends to roughly two effective radii, or $\approx$3\,kpc.

Notably, there is a mass segregation apparent in the GC population of MATLAS-2019. This is qualitatively evident in Fig.\,\ref{fig:zoom}, where we present the zoom-in of the HST observation within two effective radii. All the bright GCs are clustered towards the center of the dwarf. Quantitatively, this is shown in Fig.\,\ref{fig:radius}, where we split the density profiles into faint and bright GCs (with a magnitude cut at 23.1\,mag, which corresponds to the 1$\sigma$ of the GCLF at the bright end). The density profile of the bright GCs is centrally peaked and quickly drops to zero within 0.75\,$r_{eff}$, while the density profile of the faint GCs is more extended (up to 1.75\,$r_{eff}$) and less peaked. However, the contamination of background galaxies will increase at the faint end, so this could bias the signal. {Another caveat might be a drop in completeness towards the center of the galaxy. As discussed previously, the photometric incompleteness near the center of the galaxy does not appear to have a significant impact on source detection.} Ultimately, spectroscopic follow-up observations are needed to confirm this mass segregation.  

\section{Specific frequency} 
Is the abundance of GCs associated to MATLAS-2019 consistent with other UDGs and dwarf galaxies? One way to assess the number of GCs as a function of the brightness of the host galaxy is with the specific frequency $S_N$ \citep{1981AJ.....86.1627H}. It is calculated as:
\begin{equation}
S_N=N_{\rm GC}\cdot10^{0.4(M_V+15)}
\end{equation}
The peak of the GCLF yields a distance modulus of $m-M=31.57$\,mag. The apparent and absolute magnitudes of MATLAS-2019 are $m_V=17.69$\,mag and $M_V=-14.1\pm0.2$\,mag, respectively \citep{2021MNRAS.506.5494P}. This gives a specific frequency of $S_N=58\pm14$, where the error is a combination from the uncertainty in the absolute magnitude ($\pm0.23$\,mag) and the number of GCs ($\pm$6). 

\begin{figure*}[ht!]
\plotone{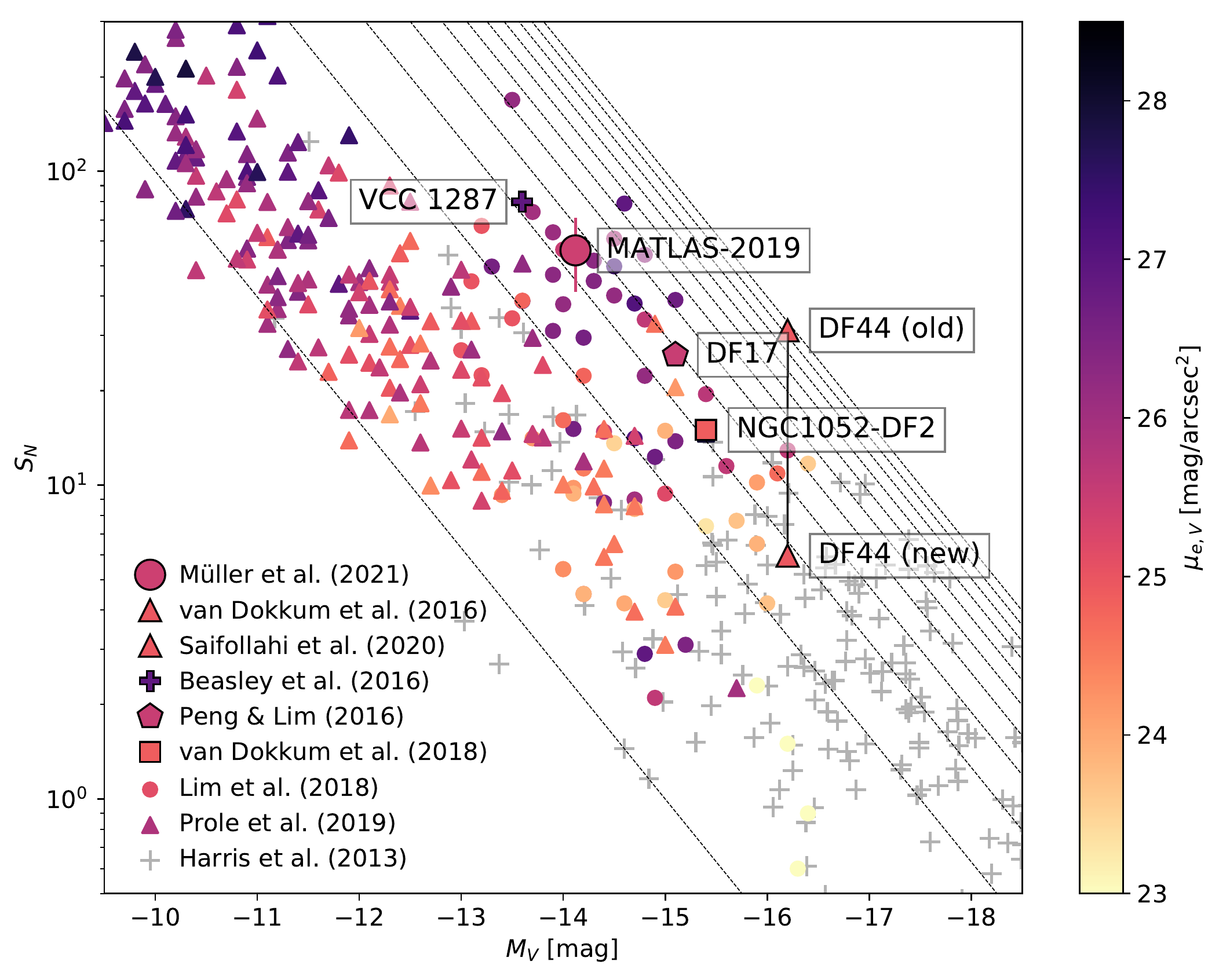}
\caption{The specific frequency of nearby galaxies \citep{2013ApJ...772...82H}, Coma cluster dwarfs \citep{2018ApJ...862...82L}, Fornax cluster dwarfs \citep{2019MNRAS.484.4865P}, and the UDGs MATLAS-2019 (this work), DF44 \citep{2020arXiv200614630S}, DF17 \citep{2016ApJ...822L..31P},  VCC1287 \citep{2016ApJ...819L..20B}, and NGC\,1052-DF2 \citep{2018ApJ...856L..30V}. The color bar indicates the effective surface brightness of the galaxies. The dashed lines correspond to specific frequencies with a constant number of GCs {of 1, 10, 20, ..., 90, 100}. \label{fig:specific}}
\end{figure*}

How does the specific frequency of MATLAS-2019 compare to other low-luminosity galaxies? In Fig.\,\ref{fig:specific} we show the specific frequency as a function of the luminosity and surface brightness of the galaxies. Several statistical studies of dwarf galaxies and UDGs \citep{2018ApJ...862...82L,2019MNRAS.484.4865P} have revealed a large scatter in the specific frequency at the faint end.  
\citet{2018ApJ...862...82L} noted that for the same luminosity, UDGs have a lower surface brightness and higher specific frequency than typical dwarf galaxies. MATLAS-2019 follows this trend. It has a high specific frequency for its luminosity and has one of the largest counts of GCs in such systems. While MATLAS-2019 doesn't stand out as having the highest specific frequency at first sight (some of the \citealt{2018ApJ...862...82L} UDGs have a higher $S_N$ value), a direct comparison between the literature UDG sample and MATLAS-2019 is tricky, mainly due to the high uncertainty of the number of GCs in the literature, as well as different approaches in counting the number of GCs. We discuss the caveats of comparing the number of GCs in the following. 

For UDGs with a high specific frequency in the \citet{2018ApJ...862...82L} UDG sample (i.e. UDGs with $S_N>30$), the median uncertainty is 33, where in the \citet{2019MNRAS.484.4865P} sample the median uncertainty is 11.  Based on the large uncertainties of those two surveys, it is expected to find some UDGs with a high number of GCs due to stochasticity. The larger error of the specific frequency estimation in these works can be explained by the fact that the dwarf galaxies studied there are either farther in the background or have been observed with ground-based facilities, where the image quality becomes an issue. In comparison to the \citet{2018ApJ...862...82L} sample, with our observations we achieve an uncertainty of 14, i.e. at a $\approx$25\% level. 

There are also different selection functions used in the literature. \citet{2018ApJ...862...82L} had to correct for their incompleteness of the GC population -- the Coma dwarfs are four times farther away than MATLAS-2019, meaning that  the full GCLF could not be traced. They selected the GCs within 1.5\,r$_{eff}$, multiplied by a factor 2 to correct for the full extent of the GC population (i.e. assuming that the half of the GC population will be within 1.5\,r$_{eff}$), and an additional factor 2 for the incompleteness of the GCLF (i.e. assuming that half the GCLF is sampled). If we estimate the number of GCs for MATLAS-2019 in the same way as \citet{2018ApJ...862...82L} we get 12 GCs within 1.5\,r$_{eff}$ and brighter than the peak of the GCLF, i.e. a total number of 48 GCs and a specific frequency of 111 for MATLAS-2019. 

\begin{figure}[ht!]
\plotone{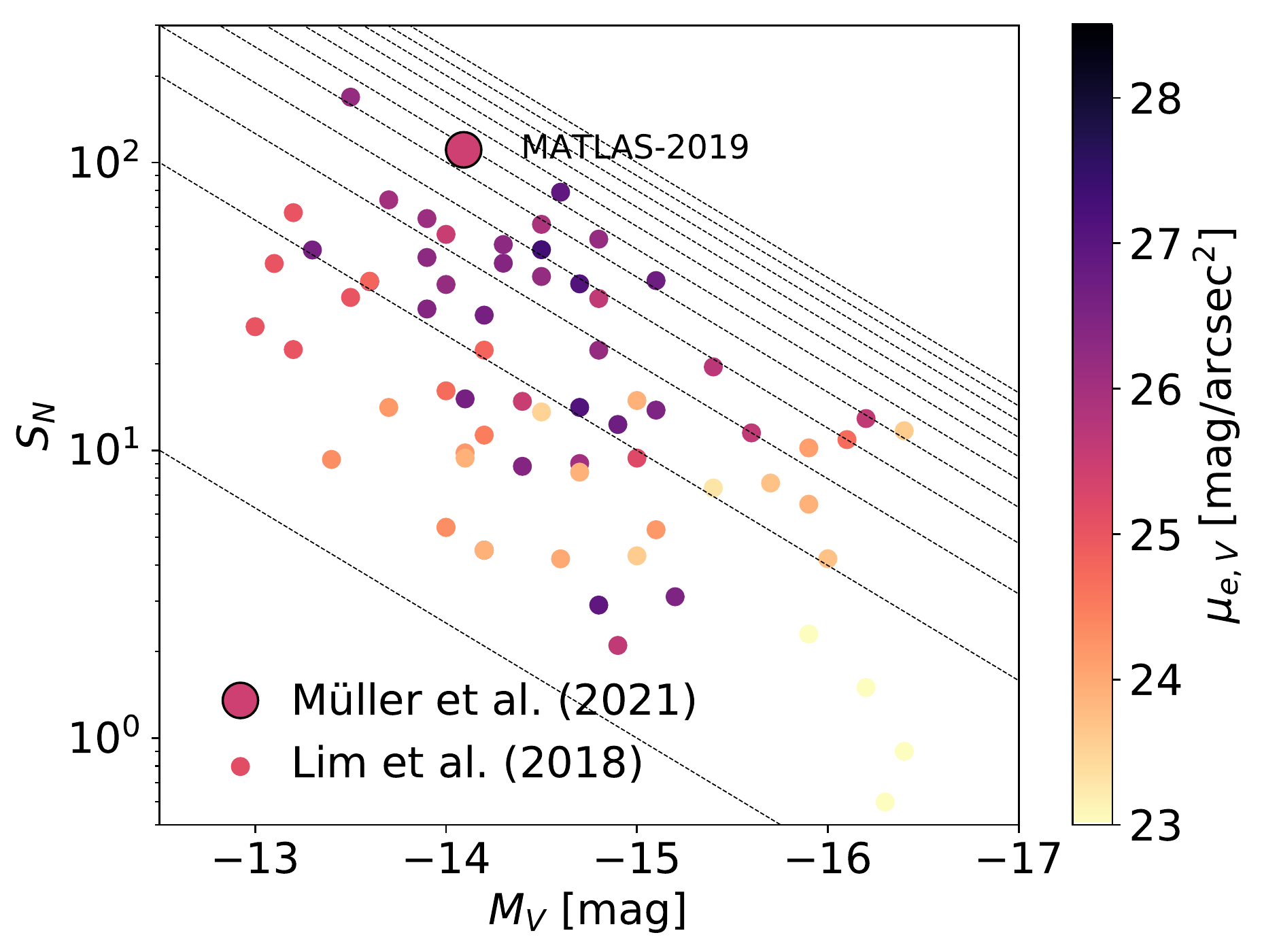}
\caption{The specific frequency of the Coma cluster dwarfs \citep{2018ApJ...862...82L} and the UDG MATLAS-2019 (this work). The specific frequency of MATLAS-2019 in this plot is estimated with the same corrections as applied in \citet{2018ApJ...862...82L} to make a direct comparison possible.  The color bar indicates the effective surface brightness of the galaxies. The dashed lines correspond to specific frequencies with a constant number of GCs, starting with one GC and then steps of 10 GCs, until 100 GCs is reached. \label{fig:specific_adapted}}
\end{figure}

In Fig.\,\ref{fig:specific_adapted} we directly compare the \citet{2018ApJ...862...82L} sample with our estimation of MATLAS-2019 adopted for their method. MATLAS-2019 stands out as the most extreme UDG.
However, the assumption that half the GCs of MATLAS-2019 are outside of 1.5\,r$_{eff}$ is not valid, as we demonstrated in Fig.\,\ref{fig:radius}. While e.g. \citet{2018MNRAS.475.4235A} found that the majority of the Coma dwarfs have $R_{GC}/R_{eff}\approx2$ (i.e., the GC system is twice as extended as the stellar light component of the galaxy), we find $R_{GC}/R_{eff}=0.7$. 
This does not mean that in general this correction is wrong, but for our particular case of MATLAS-2019 it is not appropriate, which is in line with \citet{2018MNRAS.475.4235A} finding that 4 out of their 55 dwarf galaxies have $R_{GC}/R_{eff}<1$.
For the dwarf galaxies in the Fornax cluster \citep{2019MNRAS.484.4865P}, the approach to estimate the number of GCs is similar to ours, employing Bayesian consideration by taking the background contamination into account. Therefore, the selection function should be comparable.

Other well studied UDGs with a high specific frequency are VCC\,1287, NGC\,1052-DF2, and DF44. Compared to those, MATLAS-2019 has the highest number of GCs and the highest specific frequency.
With the low uncertainty of our measurement we can claim that MATLAS-2019 is extraordinary when it comes to the abundance of GCs and possibly  {represents an} upper limit of the number of GCs in UDGs {-- at least within our current knowledge}. 
MATLAS-2019 was initially selected for follow-up studies as having the highest number of GC candidates in the MATLAS survey. This in principle suggests that we should not expect many UDGs in that sample having a higher abundance of GCs. The fact that MATLAS-2019 has such a high abundance is thus not very surprising. It also suggests that UDGs of the failed Milky Way type may be rare (at least based on environments probed by MATLAS), as we would expect them to harbour about 100 GCs or more \citep[][]{2016ApJ...828L...6V}.

\section{The dark matter mass}
The number of GCs in a system offers a means to estimate the virial mass of UDGs \citep{2016ApJ...819L..20B}. The virial mass of a galaxy scales linearly with the number of GCs over 6 orders of magnitude \citep{2013ApJ...772...82H,2020AJ....159...56B}, the relation only flattens for halos smaller than $10^{10}$\,M$_{\odot}$. According to \citet{2017ApJ...836...67H}, the virial mass $M_{\rm halo}$ of a galaxy is connected to the total mass of the GC system $ M_{\rm GC}$ with:
\begin{equation}
    M_{\rm GC}/ M_{\rm halo}=2.9\times10^{-5}
\end{equation}
Assuming a mean mass of a GC to be $1\times10^5$\,M$_{\odot}$ for dwarf galaxies \citep{2017ApJ...836...67H}, this yields a GC system mass of $M_{\rm GC}=2.6\pm0.6\times10^{6}$\,M$_{\odot}$ and
a halo mass of $M_{\rm halo}=9\pm2\times10^{10}$\,M$_{\odot}$ for MATLAS-2019. {This is above the expected range of bright dwarf galaxies  ($M_{halo}= [1.5, 5.0] \times 10^{10}$\,M$_{\odot}$, \citealt{2013ApJ...777L..10B,2017MNRAS.467.2019R,2017ARA&A..55..343B})} but also below the mass of the LMC with  $2.5\times10^{11}$\,M$_\odot$ \citep{2016MNRAS.456L..54P,2020MNRAS.495.2554E} by a factor of three, i.e. it's somewhere between the brightest dwarfs and LMC-like galaxies. With an absolute magnitude of $-14.1\pm0.2$\,mag in the $V$ band\footnote{The apparent magnitude provided in \citet{2020A&A...640A.106M} was off due to an error in the photometric calibration of the zero point. This explains the discrepancy in the photometry between \citet{2020A&A...640A.106M} and \citet{2021MNRAS.500.1279F}. The correct apparent $V$ band magnitude measured on the MATLAS data is $17.44$\,mag, which is consistent with the value of $17.48$\,mag from \citet{2021MNRAS.500.1279F}.}, this gives a mass-to-light ratio of $M_{\rm halo}/L_V=1260^{+868}_{-555}$\,M$_{\odot}$/L$_{\odot}$. The high number of GCs suggests {a rather massive halo of dark matter} for the galaxy's stellar mass, but not that of a failed Milky Way -- in comparison, the Milky Way is more than ten times more massive with a virial mass of $1.3\pm0.3\times10^{12}$\,M$_{\odot}$ estimated through its GCs \citep{2019A&A...621A..56P}.

{Finally, we present all relevant properties of MATLAS-2019 in Table \ref{tab:dwarf}.}

\begin{table}[ht]
\caption{{The photometric and the derived properties of MATLAS-2019. }
}             
\centering                          
\begin{tabular}{l c c}        
\hline\hline                 
 &  MATLAS-2019 & Reference \\
\hline      \\[-2mm]                  
RA (J2000.0) & 15:05:20.34 & (1)\\
Dec (J2000.0) & $+$01:48:44.9 & (1)\\
$m_V$ (mag) & 17.44 & (2)\\
$\mu_{eff, V}$ (mag/arcsec$^2$) & 25.6 & (2)\\
$r_{eff}$ (arcsec)& 17.2 & (2)\\
S\'ersic $n$ & 0.73 & (2) \\
$g-r$ (mag) & 0.59 & (2)\\
$g-i$ (mag) &  0.85 & (2)\\
ellipticity $e$ &0.10 & (2)\\
\\
$v_{sys}$ & 2156.0$\pm$9.4 & (3)\\
$D$ (Mpc)&  $20.7_{-2.1}^{+2.3}$ & (1)\\
$M_V$ (mag) & $-$14.1$\pm$0.2 & (1)\\
$r_{eff}$ (kpc)& 1.7$\pm$0.2& (1)\\
$\sigma_{GCs}$ (km/s) & 9.4$^{+7.0}_{-5.4}$ & (3)\\
$\sigma_{stellar}$ (km/s) & 17$\pm$2 & (4) \\
$[$Fe/H$]$ & $-$1.33$^{+0.19}_{-0.01}$ & (3)\\
age &  11.2$^{+1.8}_{-0.8}$& (3)\\
$N_{GC}$ & 26$\pm$6& (1)\\
$S_N$ & 58$\pm$14& (1) \\
$M_{\rm halo}/L_V$ ($M_\odot/L_\odot$)&$1260^{+868}_{-555}$& (1) \\
\hline
\end{tabular}
\label{tab:dwarf}
\tablecomments{The references are: (1) this work; (2) \citet{2021MNRAS.506.5494P}; (3) \citep{2020A&A...640A.106M}; and (4) \citep{2021MNRAS.500.1279F}.}
\end{table}

\section{Discussion and Conclusions}
The discovery of the ultra-diffuse galaxy MATLAS-2019 in the MATLAS deep imaging survey was followed-up with MUSE spectroscopy. While the metallicity and age estimation of the stellar body is consistent with that of other dwarf galaxies, the number of luminous GCs seemed to be too high and the measured velocity dispersion suggested a low dark matter content within this galaxy. To assess these two issues, we have observed MATLAS-2019 with the HST using two band imaging. These observations revealed a large population of 26$\pm$6 GC candidates associated with MATLAS-2019 with $V$ band magnitudes that follow a normal distribution, as expected from a well-populated GCLF. This would solve the question about the luminous GCs -- they are simply the bright end of the GCLF. We derive a specific frequency of $S_N=58\pm14$, which is at the high end of the measurements of other dwarf galaxies and may therefore  be one of the UDGs with the largest number of GCs. This is expected, since MATLAS-2019 was selected for follow-up observations because it had the highest number of possible GCs based on ground-based observations of over 2000 dwarf galaxies in the MATLAS survey. This, however, also suggests that failed Milky Way type galaxies, which were proposed as one of the origins for UDGs, may be relatively rare in the environments we have probed with MATLAS, because we would expect them to have over 100 GCs. If the UDG with the highest number of potential GCs in the MATLAS survey {has only a fraction} of the number of GCs to be considered a failed Milky Way type galaxy, it is unlikely that other UDGs in this survey are of this type.

{Because the GCs are partially resolved with the superior image quality of space-based telescopes like the HST, we were able to measure their sizes. All GCs are consistent with the sizes expected from the GCs of the Local Group.}

{
What can we learn about the history of MATLAS-2019 from the abundance of GCs? This low-surface brightness galaxy must have had high-density star formation and a high star formation rate to form such massive GCs as we observe today. And while much of the mass was locked up in GCs, the remaining stars must have spread out to form such a diffuse galaxy. \citet{2021MNRAS.tmp...71C} suggested based on simulations of cluster UDGs that they may have an excess of GCs due a combination of a) a higher star formation rate density at high redshift, when most of their star formation occurred, and b) earlier cluster infall times, shutting down the star formation. Because they would fall in earlier, they will experience more tidal effects enlarging the galaxies.
While MATLAS-2019 is not in a cluster, the NGC\,5846 group of galaxies is a dense galactic environment, so this model may provide an avenue to understand the formation history of MATLAS-2019.}

The number of GCs correlates linearly with the virial mass of a galaxy. This idea has its origins in the formation of galaxies through minor mergers: the more mergers a halo experienced, the more mass and the more GCs it accreted. Following such a trend, MATLAS-2019 with its 26 most likely GC candidates would host a vastly dark matter dominated halo with a mass-to-light ratio of over 1000. 
This provides grounds to state that MATLAS-2019 is not lacking dark matter, on the contrary, it possesses a massive dark matter halo compared to other dwarf galaxies.

\acknowledgments{
We thank the referee -- Mario Mateo -- for his excellent referee report, which greatly helped to improve this manuscript.
Based on observations with the NASA/ESA Hubble Space Telescope
obtained [from the Data Archive] at the Space Telescope Science
Institute, which is operated by the Association of Universities for
Research in Astronomy, Incorporated, under NASA contract NAS5-
26555. Support for Program number (GO-16082) was
provided through a grant from the STScI under NASA contract NAS5-
26555.  O.M. is grateful to the Swiss National Science Foundation for financial support. PRD gratefully acknowledges support from grant HST-GO-16082.002-A
M.P. acknowledges the Vice Rector for Research of the University of Innsbruck for the granted scholarship.
S.P. acknowledges support from the New Researcher Program (No. 2019R1C1C1009600) through the National Research Foundation of Korea. 
This research made use of photutils, an Astropy package for detection and photometry of astronomical sources \citep{larry_bradley_2020_4044744}.
}

\facilities{HST(STIS)}


\software{emcee \citep{2013PASP..125..306F}, \textsc{galfit}  \citep{2002AJ....124..266P},sep  \citep{SEP}, photutils  \citep{larry_bradley_2020_4044744}, astropy \citep{2013A&A...558A..33A},  Source Extractor \citep{1996A&AS..117..393B}}

\bibliography{bibliography}{}
\bibliographystyle{aasjournal}



\begin{table*}[ht]
\renewcommand{\arraystretch}{0.9}
\setlength{\tabcolsep}{1.5pt}
\small
\caption{All detected GC-like sources within 1.75 effective radius of MATLAS-2019, of which we estimate that 37 are real GC associated with MATLAS-2019. For the calculation of $r_h$ in parsec we used a distance modulus of 31.56 derived from the peak of the GCLF. The asterics denotes whether the GC was listed in \citet{2020A&A...640A.106M}. The magnitudes are extinction corrected. The errors for the absolute $V$ band magnitudes are a combination of the photometric error and the uncertainty in the distance modulus, which dominates the error.
}             
\centering                          
\begin{tabular}{l c c c c c c c c c c }        
\hline\hline                 
\# & RA  & Dec  & F606W  & F606W - F814W & $M_V$ & r$_h$  & r$_h$  & r$_c$  & r$_t$ & d$_{center}$ \\
 & (J2000.0) & (J2000.0) & (mag) & (mag) & (mag) & (arcsec) &  (pc) & (arcsec) & (arcsec) & (arcsec) \\
\hline      \\[-2mm]                  
1 & 226.328690 & 1.811024 & 24.557$\pm$0.041 & 0.739$\pm$0.050 & -6.87$\pm$0.47 & 0.058$\pm$0.004 & 5.8$\pm$0.3  &  0.025$\pm$0.004 &  0.438$\pm$0.086 & 22.5 \\
2 & 226.329288 & 1.816956 & 25.800$\pm$0.142 & 0.686$\pm$0.196 & -5.63$\pm$0.49 & 0.291$\pm$0.032 & 29.0$\pm$3.1  &  0.148$\pm$0.019 &  1.093$\pm$0.311 & 25.5 \\
3 & 226.329752 & 1.811939 & 25.147$\pm$0.073 & 0.675$\pm$0.107 & -6.28$\pm$0.48 & 0.082$\pm$0.011 & 8.2$\pm$1.1  &  0.019$\pm$0.004 &  1.111$\pm$0.584 & 18.1 \\
4$^*$ & 226.329915 & 1.811463 & 23.674$\pm$0.017 & 0.699$\pm$0.024 & -7.75$\pm$0.47 & 0.038$\pm$0.002 & 3.8$\pm$0.2  &  0.003$\pm$0.001 &  2.209$\pm$0.925 & 17.8 \\
5 & 226.330024 & 1.814198 & 24.089$\pm$0.023 & 0.717$\pm$0.029 & -7.34$\pm$0.47 & 0.021$\pm$0.003 & 2.1$\pm$0.2  &  0.008$\pm$0.003 &  0.275$\pm$0.058 & 18.1 \\
6 & 226.330036 & 1.810074 & 24.123$\pm$0.019 & 0.669$\pm$0.030 & -7.30$\pm$0.47 & 0.045$\pm$0.003 & 4.5$\pm$0.2  &  0.014$\pm$0.003 &  0.342$\pm$0.058 & 19.1 \\
7 & 226.330482 & 1.812385 & 24.475$\pm$0.038 & 0.678$\pm$0.047 & -6.95$\pm$0.47 & 0.006$\pm$0.900 & 0.6$\pm$89.7  &  0.009$\pm$0.156 &  0.034$\pm$0.385 & 15.4 \\
8 & 226.330594 & 1.812373 & 24.449$\pm$0.036 & 0.653$\pm$0.045 & -6.98$\pm$0.47 & 0.004$\pm$0.009 & 0.4$\pm$0.8  &  0.016$\pm$0.043 &  0.088$\pm$0.116 & 15.0 \\
9$^*$ & 226.331081 & 1.810577 & 24.128$\pm$0.019 & 0.650$\pm$0.030 & -7.30$\pm$0.47 & 0.062$\pm$0.004 & 6.2$\pm$0.3  &  0.015$\pm$0.002 &  0.783$\pm$0.140 & 14.9 \\
10 & 226.331100 & 1.813403 & 24.361$\pm$0.033 & 0.774$\pm$0.041 & -7.07$\pm$0.47 & 0.035$\pm$0.003 & 3.5$\pm$0.3  &  0.014$\pm$0.004 &  0.305$\pm$0.066 & 13.6 \\
11 & 226.331323 & 1.812492 & 25.081$\pm$0.071 & 0.689$\pm$0.107 & -6.35$\pm$0.48 & 0.070$\pm$0.006 & 7.0$\pm$0.6  &  0.023$\pm$0.005 &  0.573$\pm$0.195 & 12.4 \\
12$^*$ & 226.331332 & 1.815083 & 23.382$\pm$0.009 & 0.715$\pm$0.021 & -8.05$\pm$0.47 & 0.042$\pm$0.002 & 4.1$\pm$0.1  &  0.003$\pm$0.000 &  49.999$\pm$0.000 & 15.5 \\
13 & 226.331693 & 1.819393 & 25.653$\pm$0.119 & 0.795$\pm$0.166 & -5.77$\pm$0.48 & 0.130$\pm$0.028 & 12.9$\pm$2.7  &  0.043$\pm$0.013 &  0.612$\pm$0.327 & 27.3 \\
14 & 226.331837 & 1.811076 & 24.157$\pm$0.020 & 0.703$\pm$0.030 & -7.27$\pm$0.47 & 0.024$\pm$0.002 & 2.4$\pm$0.2  &  0.013$\pm$0.004 &  0.242$\pm$0.051 & 11.7 \\
15 & 226.332161 & 1.810717 & 25.613$\pm$0.117 & 0.665$\pm$0.169 & -5.81$\pm$0.48 & 0.056$\pm$0.007 & 5.6$\pm$0.7  &  0.033$\pm$0.021 &  0.240$\pm$0.124 & 11.3 \\
16 & 226.332456 & 1.808513 & 24.655$\pm$0.038 & 0.569$\pm$0.067 & -6.77$\pm$0.47 & 0.001$\pm$0.000 & 0.0$\pm$0.0  &  0.005$\pm$0.248 &  0.025$\pm$0.477 & 16.5 \\
17 & 226.332657 & 1.813898 & 25.014$\pm$0.059 & 0.586$\pm$0.100 & -6.41$\pm$0.47 & 0.026$\pm$0.005 & 2.6$\pm$0.5  &  0.015$\pm$0.016 &  0.176$\pm$0.112 & 9.2 \\
18 & 226.332700 & 1.809457 & 25.032$\pm$0.062 & 0.853$\pm$0.095 & -6.40$\pm$0.47 & 0.056$\pm$0.009 & 5.6$\pm$0.8  &  0.011$\pm$0.003 &  0.732$\pm$0.353 & 13.1 \\
19 & 226.332765 & 1.815702 & 25.654$\pm$0.119 & 0.776$\pm$0.166 & -5.77$\pm$0.48 & 0.155$\pm$0.087 & 15.5$\pm$8.7  &  0.021$\pm$0.006 &  1.010$\pm$0.802 & 13.7 \\
20 & 226.332857 & 1.812632 & 24.438$\pm$0.035 & 0.684$\pm$0.043 & -6.99$\pm$0.47 & 0.018$\pm$0.003 & 1.8$\pm$0.2  &  0.008$\pm$0.006 &  0.190$\pm$0.080 & 6.9 \\
21 & 226.333213 & 1.817496 & 25.073$\pm$0.070 & 0.639$\pm$0.107 & -6.35$\pm$0.48 & 0.024$\pm$0.004 & 2.4$\pm$0.4  &  0.018$\pm$0.024 &  0.148$\pm$0.108 & 18.9 \\
22$^*$ & 226.333478 & 1.811035 & 23.159$\pm$0.008 & 0.663$\pm$0.013 & -8.27$\pm$0.47 & 0.017$\pm$0.001 & 1.7$\pm$0.1  &  0.004$\pm$0.001 &  0.503$\pm$0.052 & 6.9 \\
23 & 226.333592 & 1.815197 & 25.063$\pm$0.068 & 0.793$\pm$0.099 & -6.36$\pm$0.47 & 0.031$\pm$0.004 & 3.1$\pm$0.4  &  0.024$\pm$0.024 &  0.156$\pm$0.093 & 10.7 \\
24 & 226.333624 & 1.815683 & 24.758$\pm$0.048 & 0.543$\pm$0.082 & -6.67$\pm$0.47 & 0.022$\pm$0.005 & 2.2$\pm$0.5  &  0.005$\pm$0.003 &  0.431$\pm$0.166 & 12.3 \\
25$^*$ & 226.333800 & 1.810589 & 23.217$\pm$0.016 & 0.723$\pm$0.019 & -8.21$\pm$0.47 & 0.013$\pm$0.001 & 1.2$\pm$0.0  &  0.006$\pm$0.003 &  0.151$\pm$0.034 & 7.6 \\
26$^*$ & 226.333895 & 1.812363 & 22.510$\pm$0.014 & 0.688$\pm$0.017 & -8.92$\pm$0.47 & 0.035$\pm$0.001 & 3.5$\pm$0.0  &  0.003$\pm$0.000 &  2.233$\pm$0.328 & 3.1 \\
27 & 226.333997 & 1.810966 & 25.012$\pm$0.059 & 0.719$\pm$0.094 & -6.42$\pm$0.47 & 0.042$\pm$0.004 & 4.1$\pm$0.4  &  0.009$\pm$0.004 &  0.568$\pm$0.254 & 6.1 \\
28 & 226.334056 & 1.810118 & 23.834$\pm$0.016 & 0.675$\pm$0.024 & -7.59$\pm$0.47 & 0.024$\pm$0.002 & 2.4$\pm$0.2  &  0.007$\pm$0.002 &  0.298$\pm$0.050 & 8.8 \\
29$^*$ & 226.334511 & 1.812922 & 22.237$\pm$0.003 & 0.695$\pm$0.006 & -9.19$\pm$0.47 & 0.060$\pm$0.001 & 6.0$\pm$0.1  &  0.005$\pm$0.000 &  2.748$\pm$0.322 & 1.9 \\
30 & 226.334833 & 1.809974 & 25.643$\pm$0.118 & 0.526$\pm$0.180 & -5.78$\pm$0.48 & 0.026$\pm$0.010 & 2.6$\pm$1.0  &  0.019$\pm$0.040 &  0.152$\pm$0.181 & 9.0 \\
31 & 226.335019 & 1.810862 & 24.548$\pm$0.040 & 0.711$\pm$0.053 & -6.88$\pm$0.47 & 0.017$\pm$0.004 & 1.7$\pm$0.4  &  0.019$\pm$0.015 &  0.134$\pm$0.059 & 5.9 \\
32$^*$ & 226.335134 & 1.813640 & 23.105$\pm$0.009 & 0.729$\pm$0.018 & -8.32$\pm$0.47 & 0.028$\pm$0.001 & 2.8$\pm$0.1  &  0.003$\pm$0.001 &  0.779$\pm$0.094 & 4.4 \\
33 & 226.335388 & 1.818917 & 26.168$\pm$0.165 & 0.807$\pm$0.226 & -5.26$\pm$0.50 & 0.001$\pm$0.422 & 0.0$\pm$42.1  &  0.006$\pm$0.928 &  0.036$\pm$3.564 & 23.3 \\
34$^*$ & 226.335522 & 1.812527 & 23.761$\pm$0.015 & 0.747$\pm$0.020 & -7.67$\pm$0.47 & 0.007$\pm$0.002 & 0.7$\pm$0.2  &  0.017$\pm$0.018 &  0.096$\pm$0.050 & 2.7 \\
35 & 226.335539 & 1.816354 & 25.390$\pm$0.107 & 0.822$\pm$0.141 & -6.04$\pm$0.48 & 0.160$\pm$0.021 & 15.9$\pm$2.0  &  0.058$\pm$0.007 &  1.264$\pm$0.512 & 14.3 \\
36$^*$ & 226.335632 & 1.811585 & 22.953$\pm$0.006 & 0.689$\pm$0.011 & -8.47$\pm$0.47 & 0.038$\pm$0.001 & 3.8$\pm$0.1  &  0.003$\pm$0.000 &  2.184$\pm$0.414 & 4.5 \\
37$^*$ & 226.335770 & 1.813566 & 23.734$\pm$0.015 & 0.717$\pm$0.020 & -7.69$\pm$0.47 & 0.001$\pm$0.220 & 0.0$\pm$21.9  &  0.010$\pm$0.065 &  0.048$\pm$0.166 & 5.4 \\
38 & 226.336036 & 1.812193 & 25.473$\pm$0.107 & 0.574$\pm$0.158 & -5.95$\pm$0.48 & 0.118$\pm$0.014 & 11.8$\pm$1.4  &  0.034$\pm$0.006 &  1.176$\pm$0.618 & 4.7 \\
39$^*$ & 226.336431 & 1.815219 & 23.318$\pm$0.009 & 0.760$\pm$0.012 & -8.11$\pm$0.47 & 0.013$\pm$0.002 & 1.3$\pm$0.2  &  0.017$\pm$0.008 &  0.117$\pm$0.029 & 11.6 \\
40$^*$ & 226.336516 & 1.817465 & 22.940$\pm$0.006 & 0.688$\pm$0.011 & -8.49$\pm$0.47 & 0.023$\pm$0.001 & 2.3$\pm$0.1  &  0.003$\pm$0.001 &  0.717$\pm$0.076 & 19.1 \\
41 & 226.336918 & 1.811995 & 25.928$\pm$0.147 & 0.653$\pm$0.210 & -5.50$\pm$0.49 & 0.210$\pm$0.149 & 20.9$\pm$14.9  &  0.026$\pm$0.007 &  4.631$\pm$14.363 & 7.9 \\
42 & 226.336924 & 1.818362 & 26.095$\pm$0.155 & 0.494$\pm$0.223 & -5.33$\pm$0.49 & 0.126$\pm$0.013 & 12.5$\pm$1.3  &  0.361$\pm$0.918 &  0.275$\pm$0.092 & 22.6 \\
43 & 226.337516 & 1.817296 & 24.386$\pm$0.034 & 0.616$\pm$0.043 & -7.04$\pm$0.47 & 0.028$\pm$0.003 & 2.8$\pm$0.3  &  0.005$\pm$0.002 &  0.528$\pm$0.142 & 20.0 \\
44 & 226.338322 & 1.818318 & 24.729$\pm$0.045 & 0.652$\pm$0.070 & -6.70$\pm$0.47 & 0.049$\pm$0.005 & 4.8$\pm$0.5  &  0.008$\pm$0.002 &  0.721$\pm$0.266 & 24.6 \\
45 & 226.340142 & 1.817091 & 24.735$\pm$0.046 & 0.514$\pm$0.081 & -6.69$\pm$0.47 & 0.001$\pm$0.317 & 0.0$\pm$31.5  &  0.008$\pm$0.269 &  0.036$\pm$0.816 & 25.5 \\
46 & 226.340199 & 1.810136 & 25.933$\pm$0.147 & 0.767$\pm$0.202 & -5.49$\pm$0.49 & 0.120$\pm$0.014 & 12.0$\pm$1.4  &  0.054$\pm$0.010 &  1.423$\pm$0.990 & 21.3 \\
47 & 226.340572 & 1.809045 & 24.764$\pm$0.049 & 0.743$\pm$0.069 & -6.66$\pm$0.47 & 0.075$\pm$0.005 & 7.5$\pm$0.5  &  0.017$\pm$0.003 &  0.772$\pm$0.229 & 24.3 \\
48 & 226.340587 & 1.814754 & 25.803$\pm$0.143 & 0.817$\pm$0.188 & -5.62$\pm$0.49 & 0.001$\pm$0.000 & 0.0$\pm$0.0  &  0.002$\pm$0.000 &  49.999$\pm$0.000 & 22.5 \\
49 & 226.341214 & 1.811742 & 25.313$\pm$0.098 & 0.713$\pm$0.138 & -6.12$\pm$0.48 & 0.114$\pm$0.008 & 11.3$\pm$0.8  &  0.043$\pm$0.007 &  0.600$\pm$0.173 & 23.4 \\
\hline
\end{tabular}
\label{tab:sample}
\end{table*}

\end{document}